\documentclass{PoS}
\pdfoutput=1

\usepackage[pdftex]{graphicx}
\usepackage{subfig}
\usepackage{float}
\usepackage{url}
\usepackage{braket}
\newcommand{\Tr}{\mbox{\rm Tr}}
\newcommand{\ReC}{\mbox{\rm Re}}

\def\NPB{{\em Nucl. Phys.} B}
\def\PLB{{\em Phys. Lett.}  B}

\def\PRD{{\em Phys. Rev.} D}
\def\PRC{{\em Phys. Rev.} C}

\def\be{\begin{equation}}
\def\ee{\end{equation}}
\def\bea{\begin{eqnarray}}
\def\eea{\end{eqnarray}}

\usepackage{graphicx} 

\newcommand{\trD}[1]{\mbox{\boldmath $#1$}}

\newcommand{\vers}[1]{\hat{\trD{#1}}}

\newcommand{\spub}[2]{\overline{u}_{#1}(#2)}
\newcommand{\spv}[2]{v_{#1}(#2)}

\title{Schwinger-Dyson equations and the quark-antiquark static potential}

\ShortTitle{Schwinger-Dyson equations and the quark-antiquark static potential}

\author{\speaker{Pedro Bicudo}
\\ CFTP, Instituto Superior T\'ecnico, Lisboa 
\\ E-mail: \email{bicudo@ist.utl.pt}}
\author{Gon\c{c}alo Marques, Marco Cardoso, Nuno Cardoso\\
\\ CFTP, Instituto Superior T\'ecnico, Lisboa 
}
\author{Orlando Oliveira\\
\\ CFC, Departamento de F'{\i}sica, Universidade de Coimbra, 
}

\abstract{
In lattice QCD, a confining potential for a static quark-antiquark pair can be computed with the Wilson loop technique. This potential, dominated by a linear potential at moderate distances, is consistent with the confinement with a flux tube, an extended and scalar system also directly observable in lattice QCD. Quantized flux tubes have also been observed in another class of confinement, the magnetic confinement in type II superconductors. On the other hand the solution of Schwinger Dyson Equations, say with the Landau gauge fixing and the truncation of the series of Feynman diagrams, already at the rainbow level for the self energy and at the ladder level for the Bethe Salpeter equation, provides a signal of a possible inverse quartic potential in momentum space derived from one gluon and one ghost exchange, consistent with confinement. Here we address the successes, difficulties and open problems of the matching of these two different perspectives of confinement, the Schwinger-Dyson perspective versus the flux tube perspective.
}

\FullConference{
International Workshop on QCD Green's Functions, Confinement and Phenomenology
\\ September 7-11 , 2009
\\ ECT Trento, Italy}

\begin{document}

\section{Introduction}

Bjorken
\cite{Bjorken}
asked, ``how are the many disparate methods of 
describing hadrons which are now in use related to each other 
and to the first principles of QCD?''.
This is an important question, since the non-pertubative nature of QCD leads to different truncation schemes or models,
which turn out to be difficult to match.
Apparently chiral symmetry breaking and scalar confinement 
are conflicting,
because chiral symmetry breaking requires a chiral invariant
coupling to the quarks, say a vector coupling like in QCD.
Here we show a possible solution to this old conflict of hadronic physics, 
which remained open for many years.

\begin{figure}[H]
\begin{centering}
    \subfloat[\label{fig:shapeU}U shape geometry.]{
\begin{centering}
    \includegraphics[height=4cm]{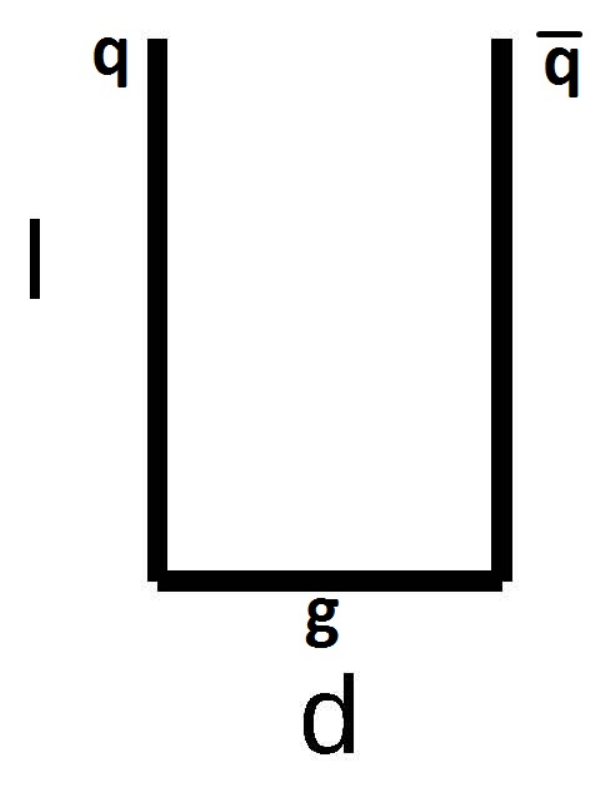}
\par\end{centering}}
    \subfloat[\label{fig:shapeL}L shape geometry.]{
\begin{centering}
    \includegraphics[height=4cm]{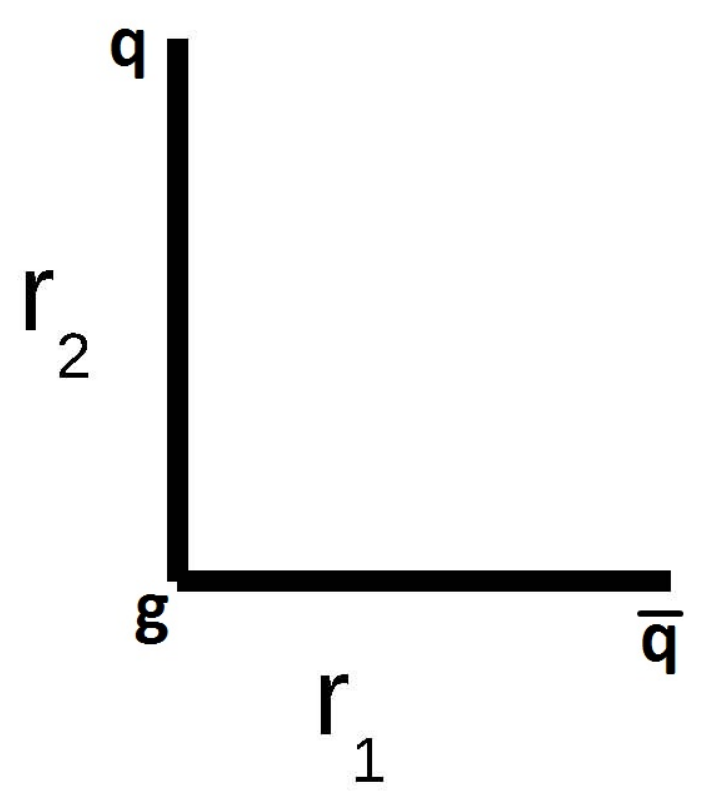}
\par\end{centering}}
\par\end{centering}
    \caption{gluon-quark-antiquark geometries, U and L shapes.}
    \label{shape}
\end{figure}

\begin{figure}[h]
\begin{center}
    \includegraphics[width=7cm]{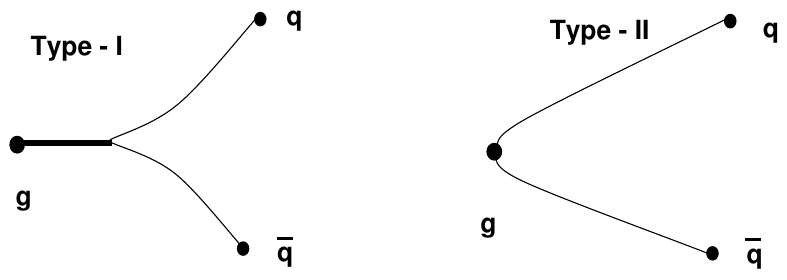}
    \caption{String attraction and fusion, and string repulsion, respectively in type I and II superconductors.}
    \label{superconductor}
\end{center}
\end{figure}

This talk has two goals.  Our first aim is to compute in Lattice QCD a confining flux tube, to illustrate scalar confinement. Because the meeting QCD-TNT is partly focused on gluons, and because the flux tubes for mesons and for baryons have been already studied in the literature, here we study the novel case of the hybrid gluon-quark-antiquark system. 
In Bicudo et al. \cite{Bicudo2008a} and Cardoso et al. \cite{Cardoso2007} it was shown that
for the static gluon-quark-antiquark potential $V$ when the segments between gluon-quark and gluon-antiquark are perpendicular, $V$ is compatible with confinement realized with a pair of fundamental strings, one linking the gluon to the quark and the other linking the same gluon to the antiquark.
However, if the segments are parallel and superposed, the total string tension becomes larger and agrees with Casimir Scaling measured by Bali \cite{Bali2000}. This can be interpreted with a type-II superconductor analogy for the confinement in QCD, see figure \ref{superconductor}, with repulsion of the fundamental strings and with the string tension of the first topological excitation of the string (the adjoint string) larger than the double of the fundamental string tension.
Here we go beyond this study of the static potentials, we compute 
the chromoelectric and chromomagnetic fields and study the shape of the flux-tubes .
We study two different geometries for the hybrid system, one with a U shape and another with L shape, figure. \ref{shape}.

Our second aim is to show an example of a quark model for a possible coupling of a scalar string to quarks, that matches the apparently conflicting vector 
coupling of QCD with a scalar confinement. 
The QCD Lagrangian is chiral invariant in the 
limit of vanishing quark masses, 
because the coupling of a gluon to a quark is vector like,
and it is this vector coupling that apparently conflicts with
scalar confinement. 
We also solve the mass gap equation for the 
dynamical generation of the quark mass with the spontaneous breaking of chiral symmetry,
and we indeed generate both the constituent quark mass and the scalar confinement.

\section{The Wilson Loops and Chromo Fields for exotic systems}
\begin{figure}[h]
\begin{centering}
    \subfloat[\label{loop0}]{
\begin{centering}
    \includegraphics[width=5cm]{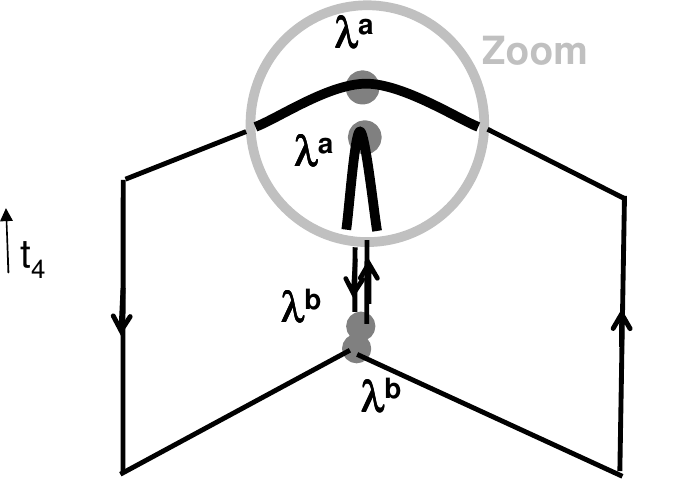}
\par\end{centering}}
    \subfloat[\label{loop1}]{
\begin{centering}
    \includegraphics[width=7cm]{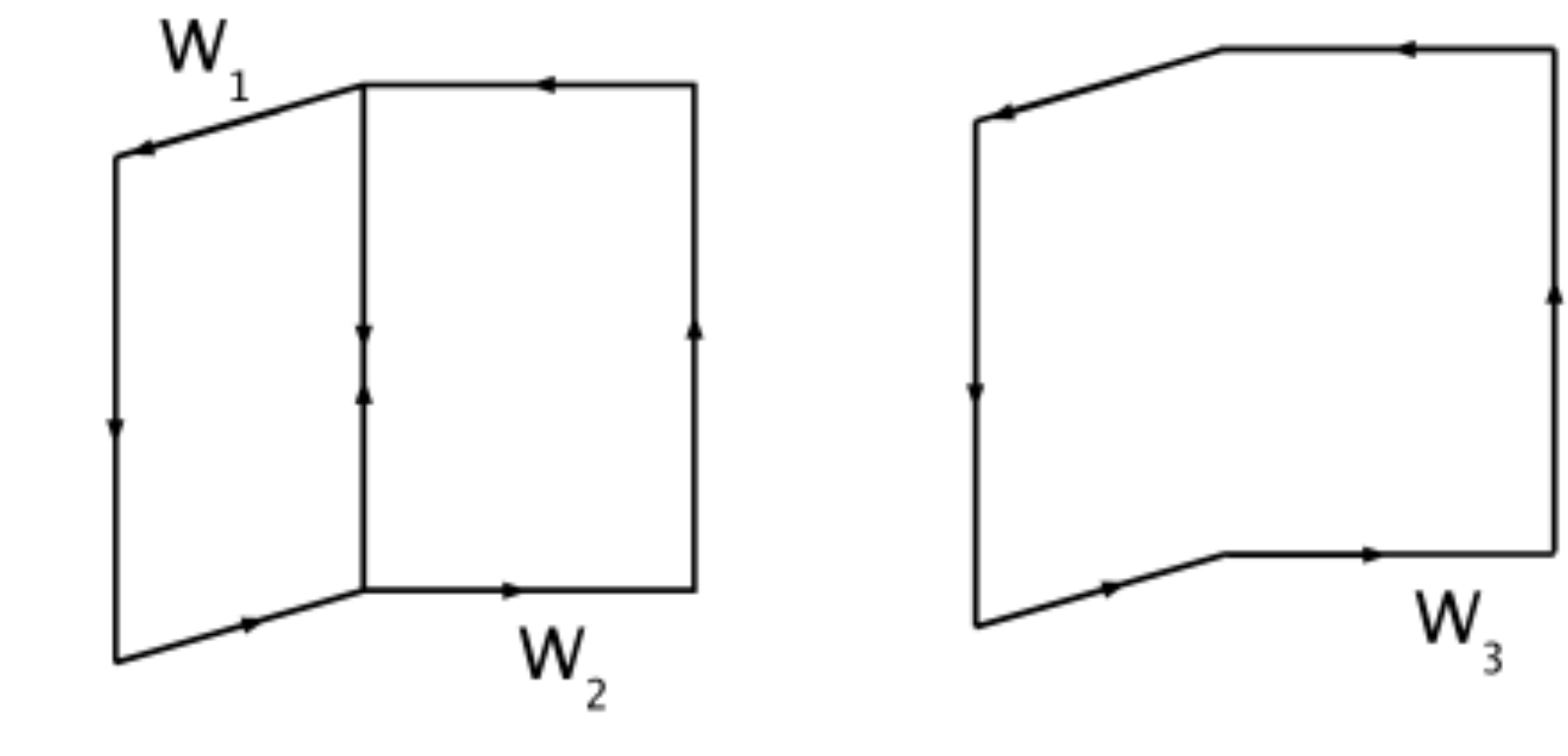}
\par\end{centering}}
\par\end{centering}
    \caption{\protect\subref{loop0} Wilson loop for the $gq\overline{q}$ and equivalent position of the static antiquark, gluon, and quark. \protect\subref{loop1} Simple Wilson loops that make the $gq\overline{q}$ Wilson loop.}
    \label{loop}
\end{figure}

In principle, any Wilson loop with a geometry similar to that represented in figure \ref{loop0}, describing correctly the quantum numbers of the hybrid, is appropriate, although the signal to noise ratio may depend on the choice of the Wilson loop.
A correct Wilson loop must include an SU(3) octet (the gluon), an SU(3) triplet (the quark) and an SU(3) antitriplet (the antiquark), as well as the connection between the three links of the gluon, the quark and the antiquark.

The Wilson loop for this system is given by:
\begin{eqnarray}
    W_{gq\overline{q}} & = & \frac{1}{16}Tr[\lambda^b T^+ \lambda^a T]Tr[\lambda^b S_1 \lambda^a S_2]\\
    & = & W_1 W_2 - \frac{1}{3}W_3
\end{eqnarray}
where $W_1$, $W_2$ and $W_3$ are the simple Wilson loops shown in figure \ref{loop1}.
When $r_1=0$, $W_1=3$ and $W_2=W_3$, the operator reduces to the mesonic Wilson loop
and when $\mu=\nu$ and $r_1=r_2=r$, $W_2=W_1^{\dagger}$ and $W_3=3$, $W_{gq\overline{q}}$ reduces
to $W_{gq\overline{q}}(r,r,t)=|W(r,t)|^2-1$, that is the Wilson loop in the adjoint representation
used to compute the potential between two static gluons.

In order to improve the signal to noise ratio of the Wilson loop, the links are replaced by "fat links",
\begin{eqnarray}
    U_{\mu}\left(s\right) & \rightarrow & P_{SU(3)}\frac{1}{1+6w}\left(U_{\mu}\left(s\right) +w\sum_{\mu\neq\nu}U_{\nu}\left(s\right)U_{\mu}\left(s+\nu\right)U_{\nu}^{\dagger}\left(s+\mu\right)\right) \ ,
\end{eqnarray}
using $w = 0.2$ and iterating this procedure 25 times in the spatial direction.

We ontain the chromoelectric and chromomagnetic fields with,
\begin{equation}
    \Braket{E^2}= \Braket{P_{0i}}-\frac{\Braket{W\,P_{0i}}}{\Braket{W}}
\end{equation}
\begin{equation}
    \Braket{B^2}= \frac{\Braket{W\,P_{ij}}}{\Braket{W}}-\Braket{P_{ij}}
\end{equation}
and the plaquette is given by
\begin{equation}
P_{\mu\nu}\left(s\right)=1 - \frac{1}{3} \ReC\Tr\left[ U_{\mu}(s) U_{\nu}(s+\mu) U_{\mu}^\dagger(s+\nu) U_{\nu}^\dagger(s) \right]
\end{equation}
We only apply the smearing technique to the Wilson loop.
The relationship between various plaquettes and the field components is:
\begin{center}
\begin{tabular}{cc}
component & plaquette\tabularnewline
\hline
 $E_x$ & $P_{xt}$ \tabularnewline
 $E_y$ & $P_{yt}$ \tabularnewline
 $E_z$ & $P_{zt}$ \tabularnewline
 $B_x$ & $P_{yz}$ \tabularnewline
 $B_y$ & $P_{zx}$ \tabularnewline
 $B_z$ & $P_{xy}$ \tabularnewline
\end{tabular}
\end{center}

The energy ($\varepsilon$) and action ($\gamma$) density is given by
\begin{equation}
    \varepsilon = \frac{1}{2}\left( \Braket{E^2} + \Braket{B^2}\right)
\end{equation}
\begin{equation}
    \gamma = \frac{1}{2}\left( \Braket{E^2} - \Braket{B^2}\right)
\end{equation}

\section{Results for the fluxt tube created by a gluon, a quark and an antiquark}
Here we present the results of our simulations with 141 $24^3 \times 48$, $\beta = 6.2$ configurations generated with the version 6 of the MILC code \cite{MILC}, via a combination of Cabbibo-Mariani and overrelaxed updates. The results are presented in lattice spacing units
$a = 0.072 $ fm.

\begin{figure}[H]
\begin{centering}
    \subfloat[Chromoelectric Field\label{fig:cfield_U_d_0_l_8_E}]{
\begin{centering}
    \includegraphics[width=3.5cm]{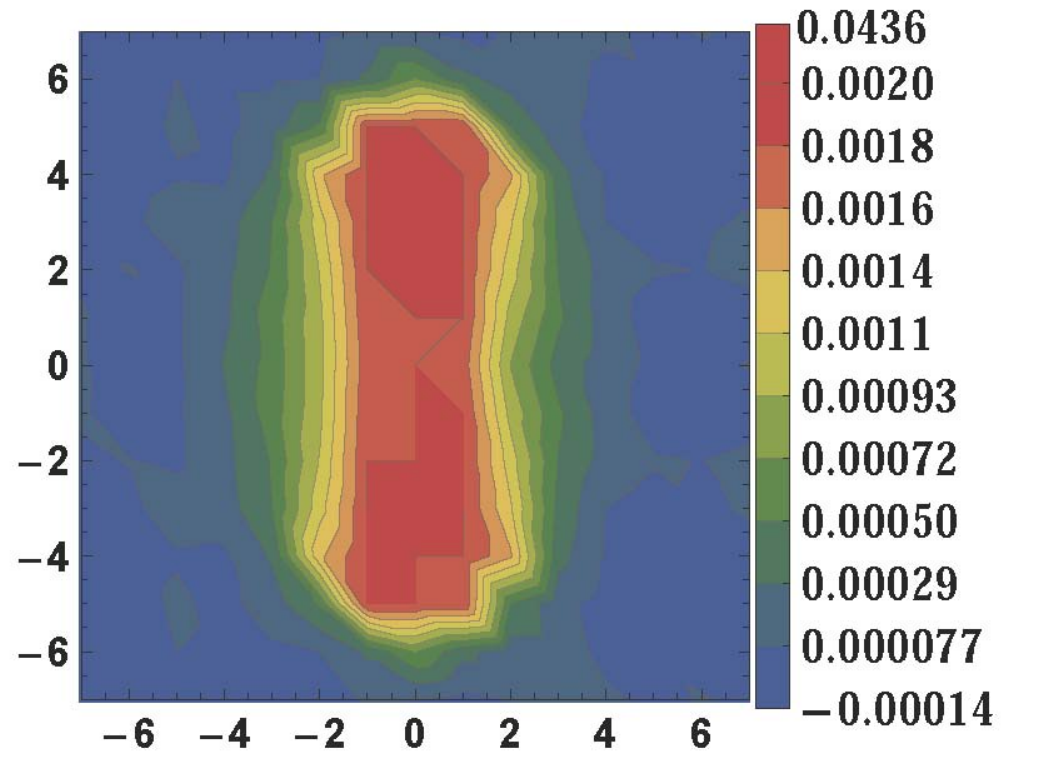}
\par\end{centering}}
    \subfloat[Chromomagnetic Field\label{fig:cfield_U_d_0_l_8_B}]{
\begin{centering}
    \includegraphics[width=3.5cm]{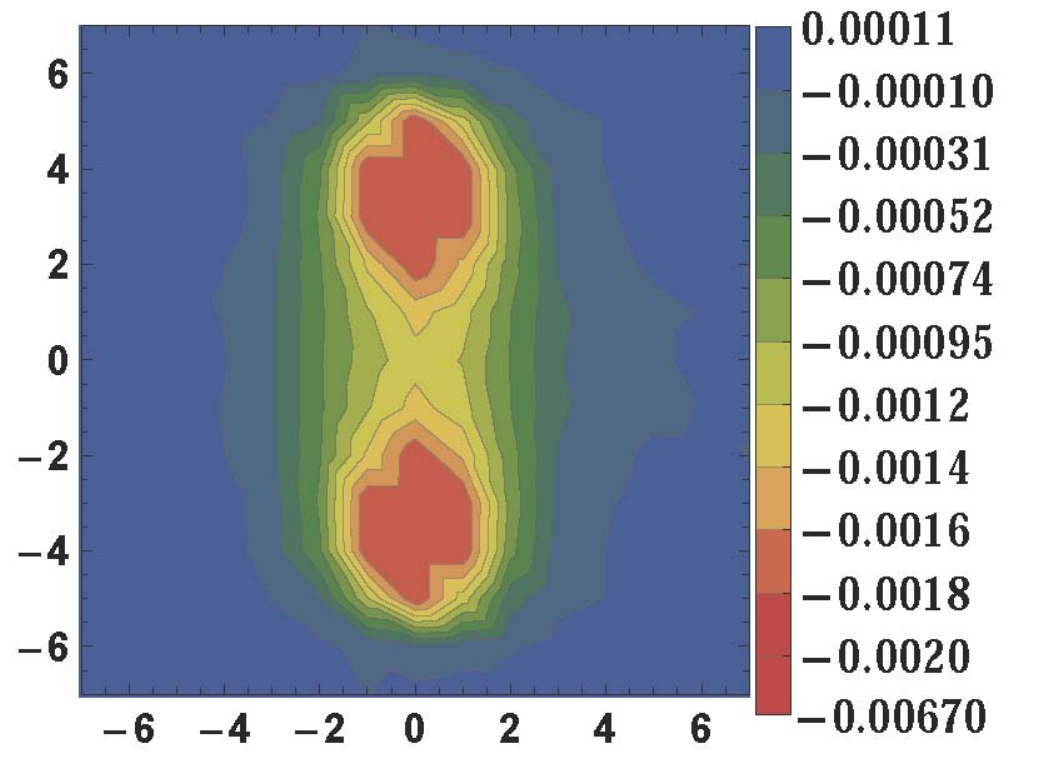}
\par\end{centering}}
    \subfloat[Energy Density\label{ig:cfield_U_d_0_l_8_Energ}]{
\begin{centering}
    \includegraphics[width=3.5cm]{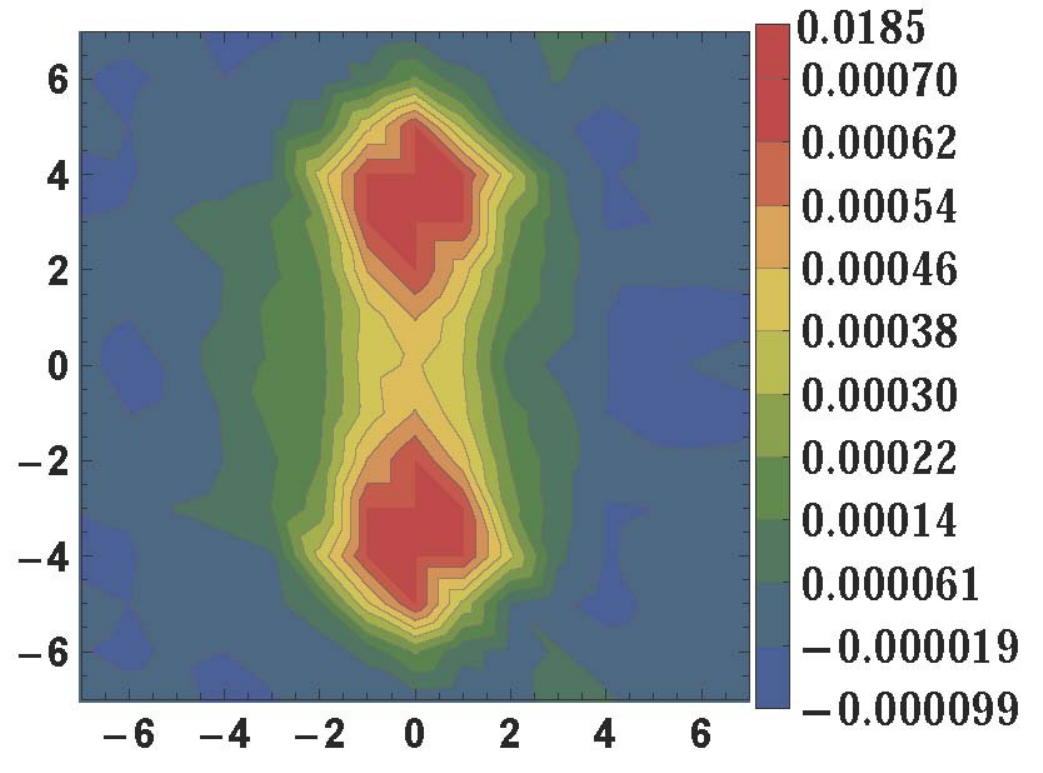}
\par\end{centering}}
    \subfloat[Action Density\label{fig:cfield_U_d_0_l_8_Act}]{
\begin{centering}
    \includegraphics[width=3.5cm]{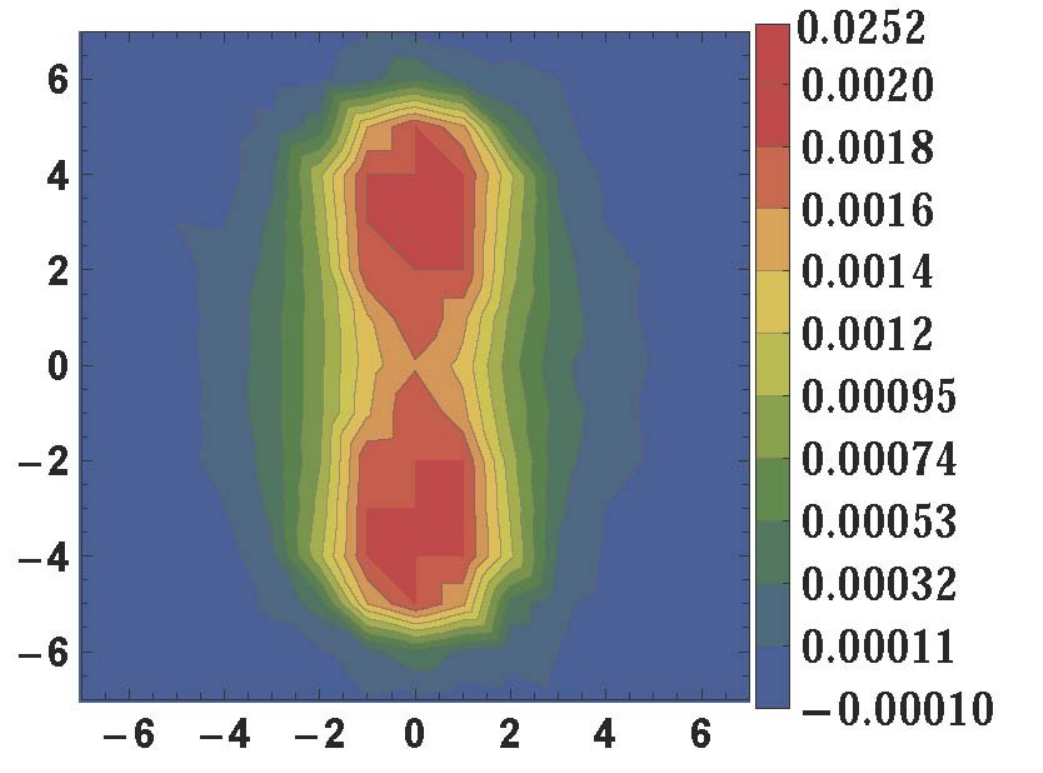}
\par\end{centering}}
\par\end{centering}
    \caption{Results for the U geometry with $d=0$ and $l=8$. In this case, the quark and the antiquark are superposed and the results are consistent
with the degenerate case of the two gluon glueball.}
    \label{latfig/cfield_U_d_0_l_8}
\end{figure}

\begin{figure}[H]
\begin{centering}
    \subfloat[Chromoelectric Field\label{fig:cfield_U_d_2_l_8_E}]{
\begin{centering}
    \includegraphics[width=3.5cm]{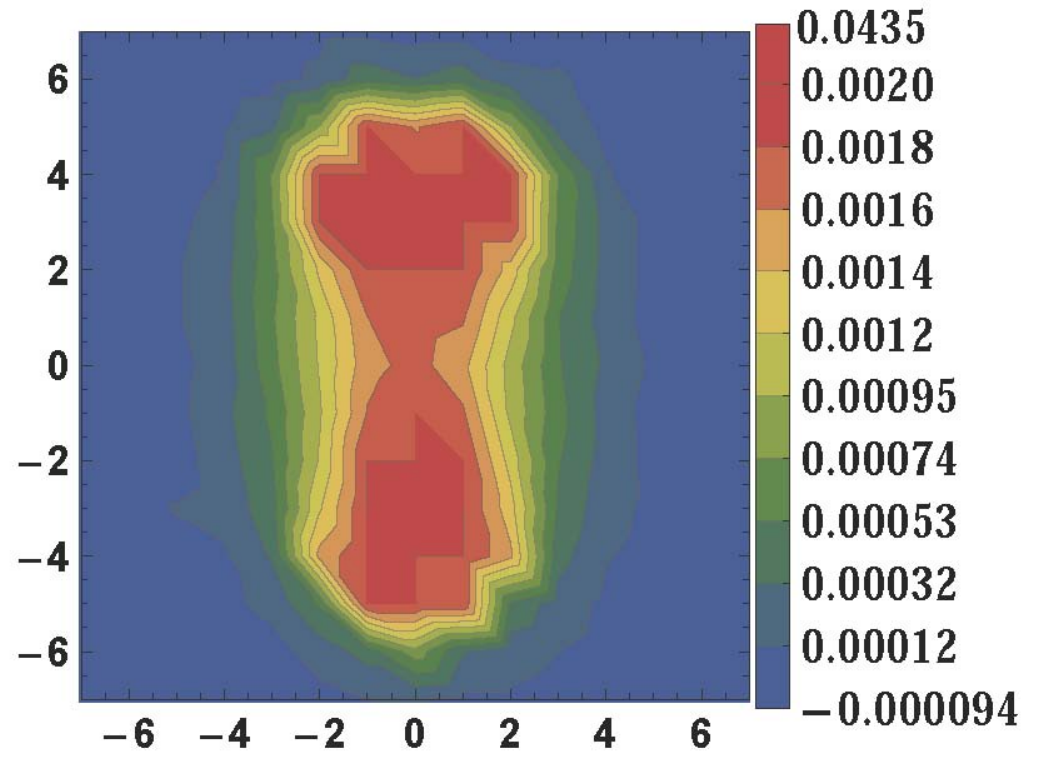}
\par\end{centering}}
    \subfloat[Chromomagnetic Field\label{fig:cfield_U_d_2_l_8_B}]{
\begin{centering}
    \includegraphics[width=3.5cm]{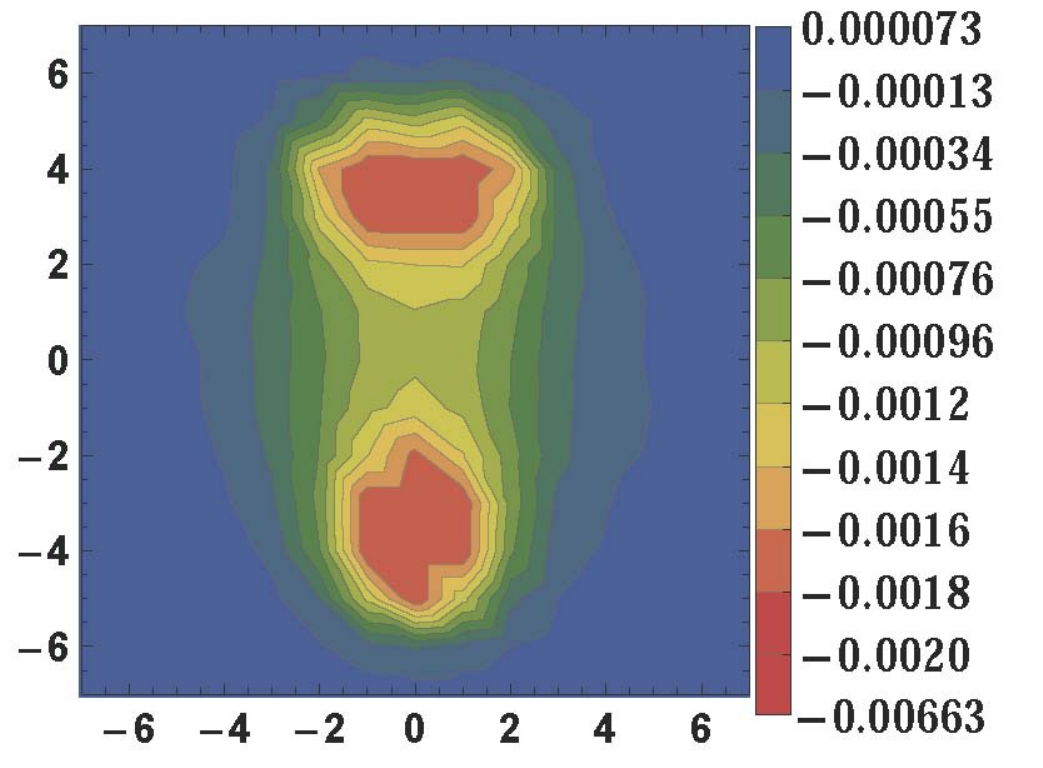}
\par\end{centering}}
    \subfloat[Energy Density\label{fig:cfield_U_d_2_l_8_Energ}]{
\begin{centering}
    \includegraphics[width=3.5cm]{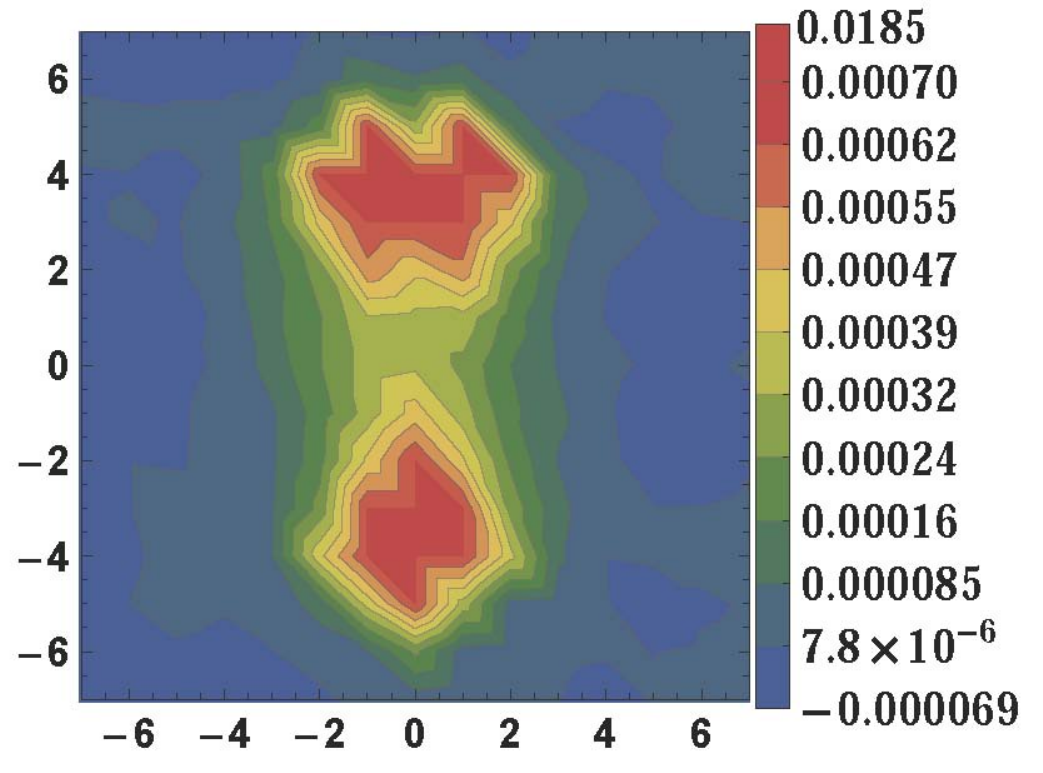}
\par\end{centering}}
    \subfloat[Action Density\label{fig:cfield_U_d_2_l_8_Act}]{
\begin{centering}
    \includegraphics[width=3.5cm]{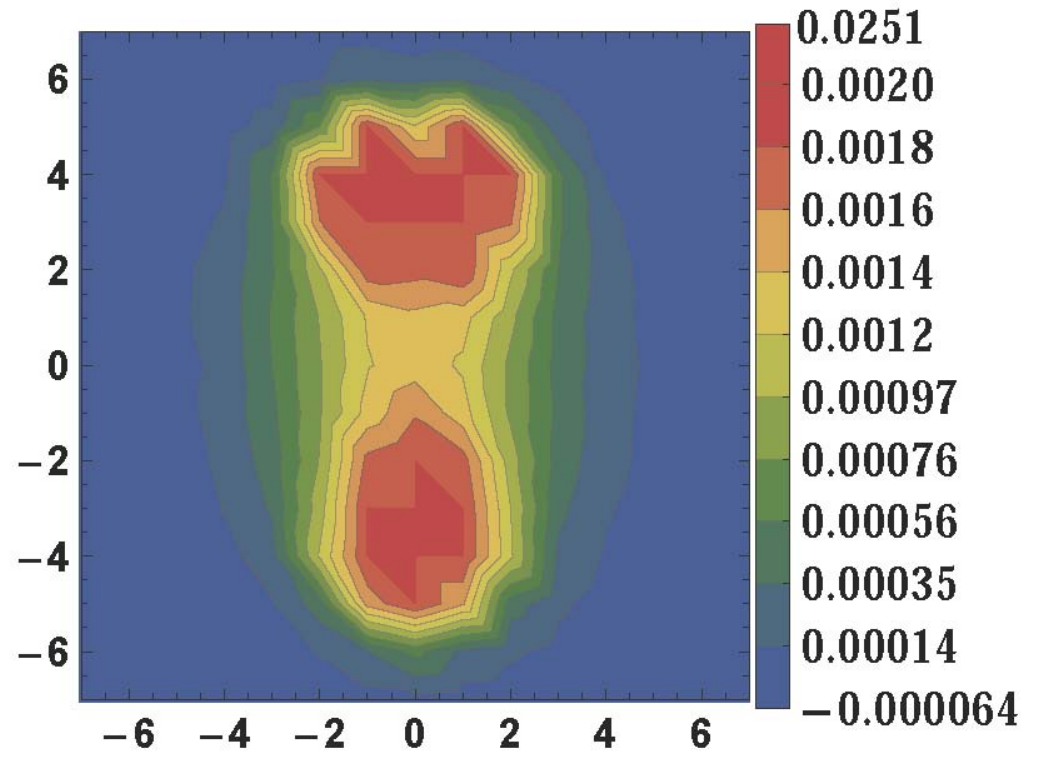}
\par\end{centering}}
\par\end{centering}
    \caption{Results for the U geometry with $d=2$ and $l=8$}
    \label{latfig/cfield_U_d_2_l_8}
\end{figure}

\begin{figure}[H]
\begin{centering}
    \subfloat[Chromoelectric Field\label{fig:cfield_U_d_4_l_8_E}]{
\begin{centering}
    \includegraphics[width=3.5cm]{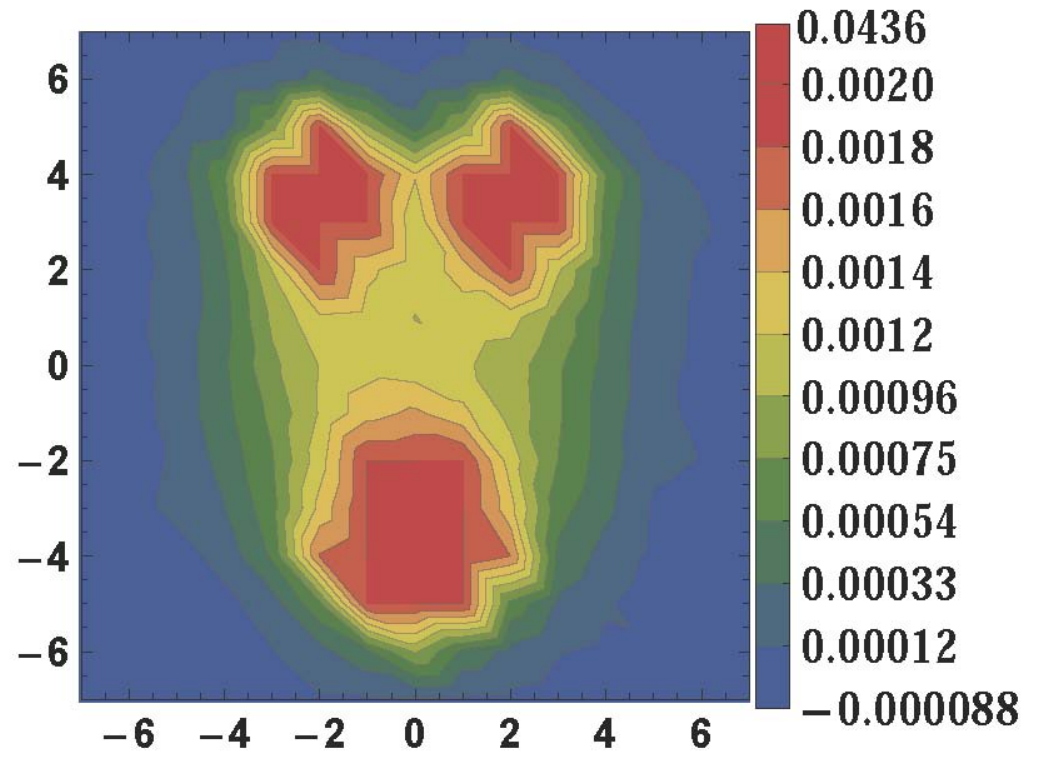}
\par\end{centering}}
    \subfloat[Chromomagnetic Field\label{fig:cfield_U_d_4_l_8_B}]{
\begin{centering}
    \includegraphics[width=3.5cm]{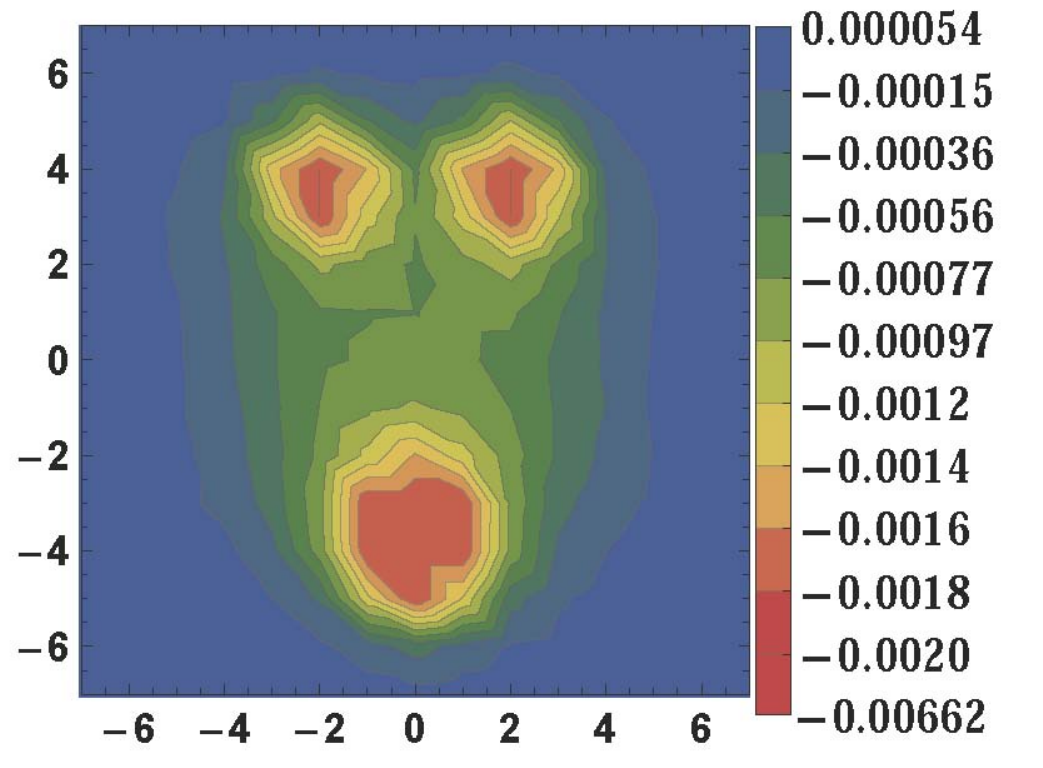}
\par\end{centering}}
    \subfloat[Energy Density\label{fig:cfield_U_d_4_l_8_Energ}]{
\begin{centering}
    \includegraphics[width=3.5cm]{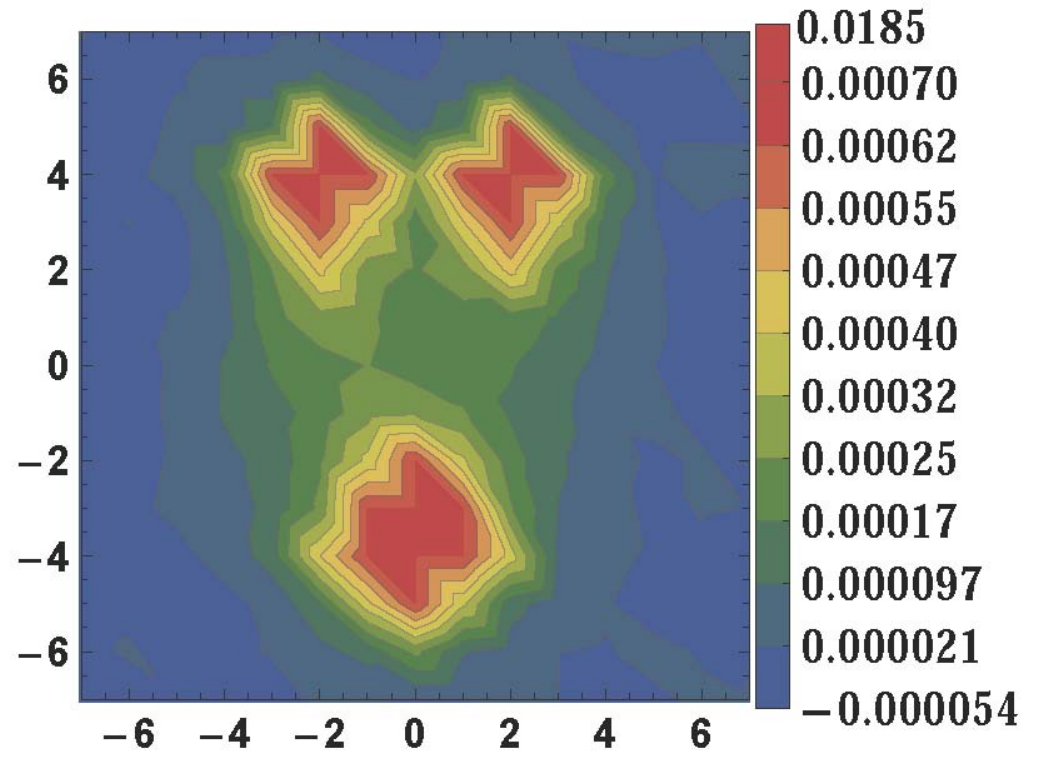}
\par\end{centering}}
    \subfloat[Action Density\label{fig:cfield_U_d_4_l_8_Act}]{
\begin{centering}
    \includegraphics[width=3.5cm]{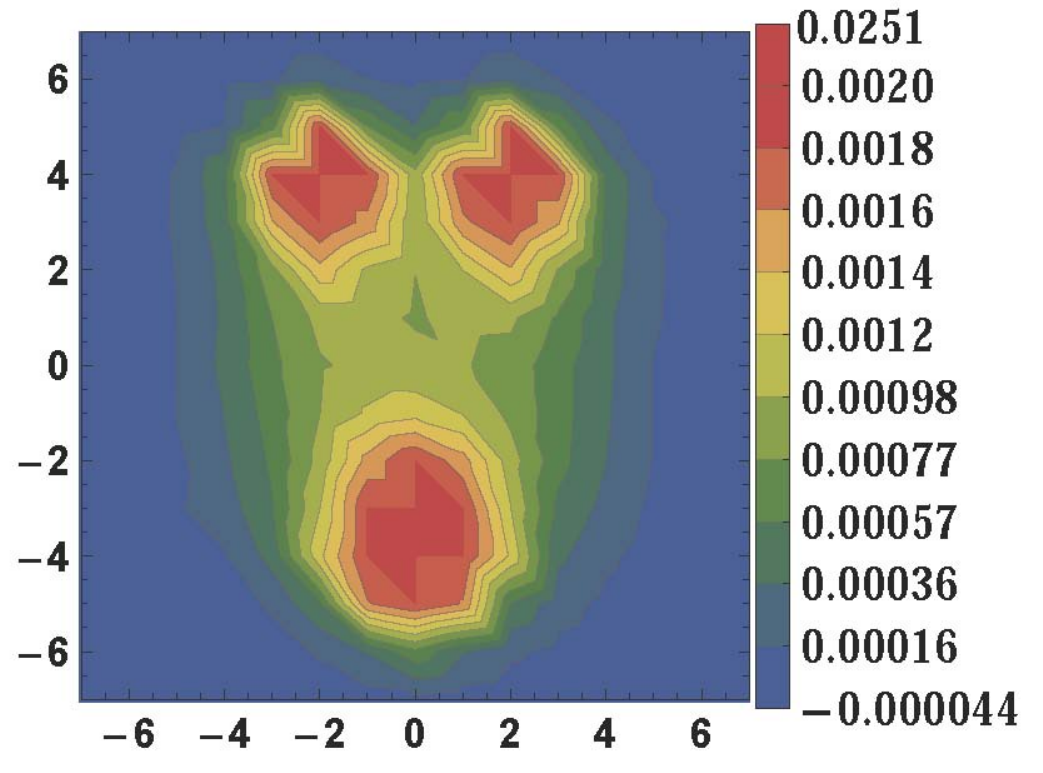}
\par\end{centering}}
\par\end{centering}
    \caption{Results for the U geometry with $d=4$ and $l=8$}
    \label{latfig/cfield_U_d_4_l_8}
\end{figure}

\begin{figure}[H]
\begin{centering}
    \subfloat[Chromoelectric Field\label{fig:cfield_U_d_6_l_8_E}]{
\begin{centering}
    \includegraphics[width=3.5cm]{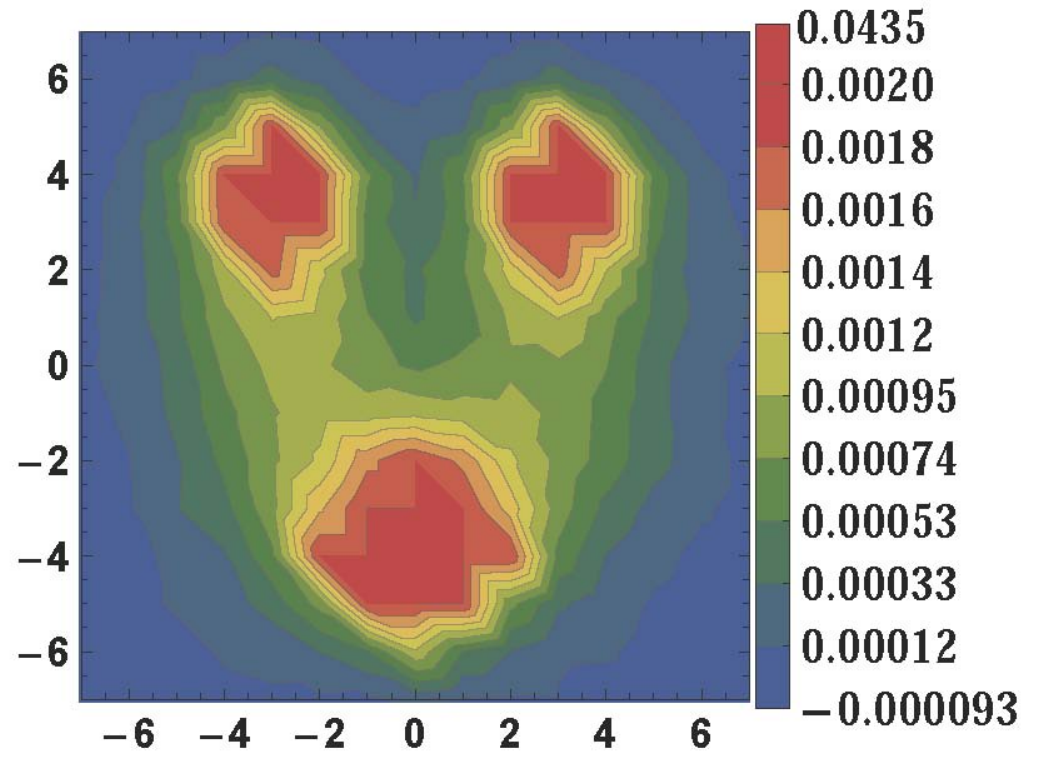}
\par\end{centering}}
    \subfloat[Chromomagnetic Field\label{fig:cfield_U_d_6_l_8_B}]{
\begin{centering}
    \includegraphics[width=3.5cm]{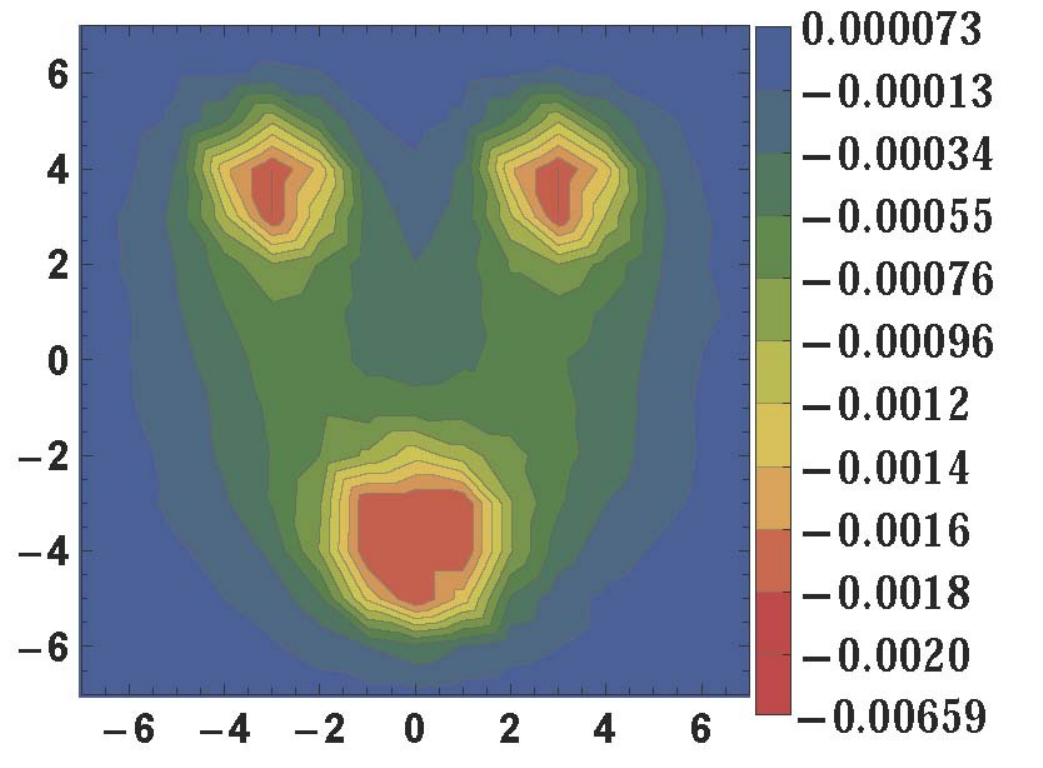}
\par\end{centering}}
    \subfloat[Energy Density\label{fig:cfield_U_d_6_l_8_Energ}]{
\begin{centering}
    \includegraphics[width=3.5cm]{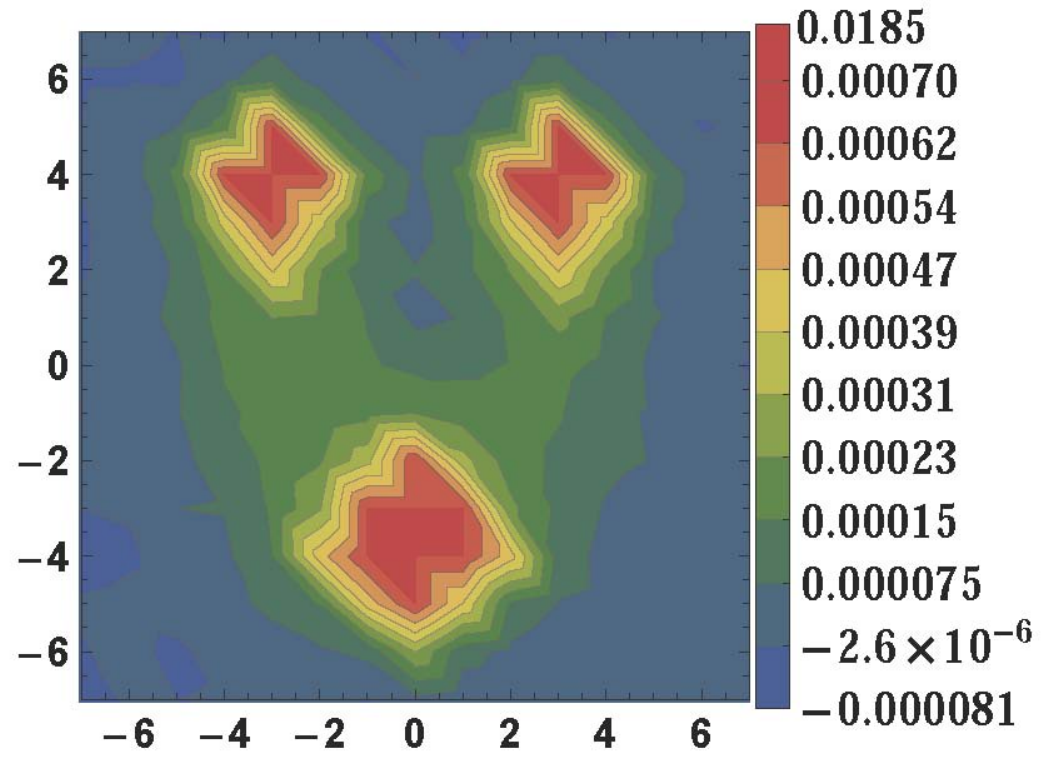}
\par\end{centering}}
    \subfloat[Action Density\label{fig:cfield_U_d_6_l_8_Act}]{
\begin{centering}
    \includegraphics[width=3.5cm]{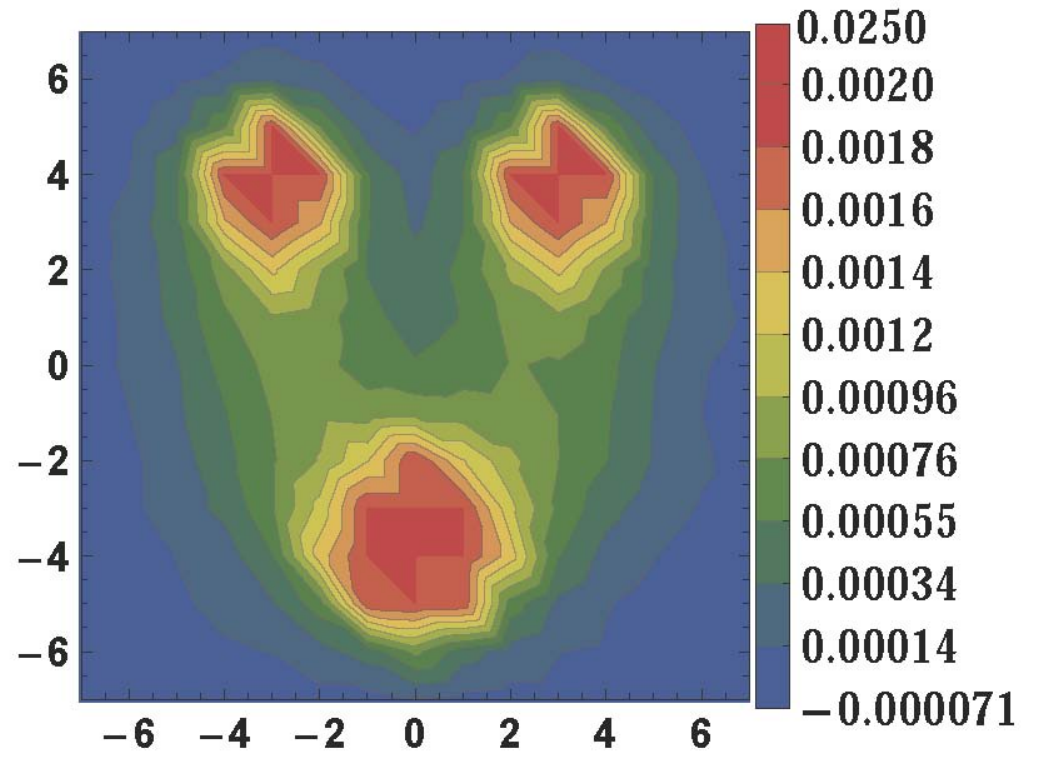}
\par\end{centering}}
\par\end{centering}
    \caption{Results for the U geometry with $d=6$ and $l=8$}
    \label{latfig/cfield_U_d_6_l_8}
\end{figure}

\begin{figure}[H]
\begin{centering}
    \subfloat[Chromoelectric Field\label{fig:cfield_L_r1_0_r2_8_E}]{
\begin{centering}
    \includegraphics[width=3.5cm]{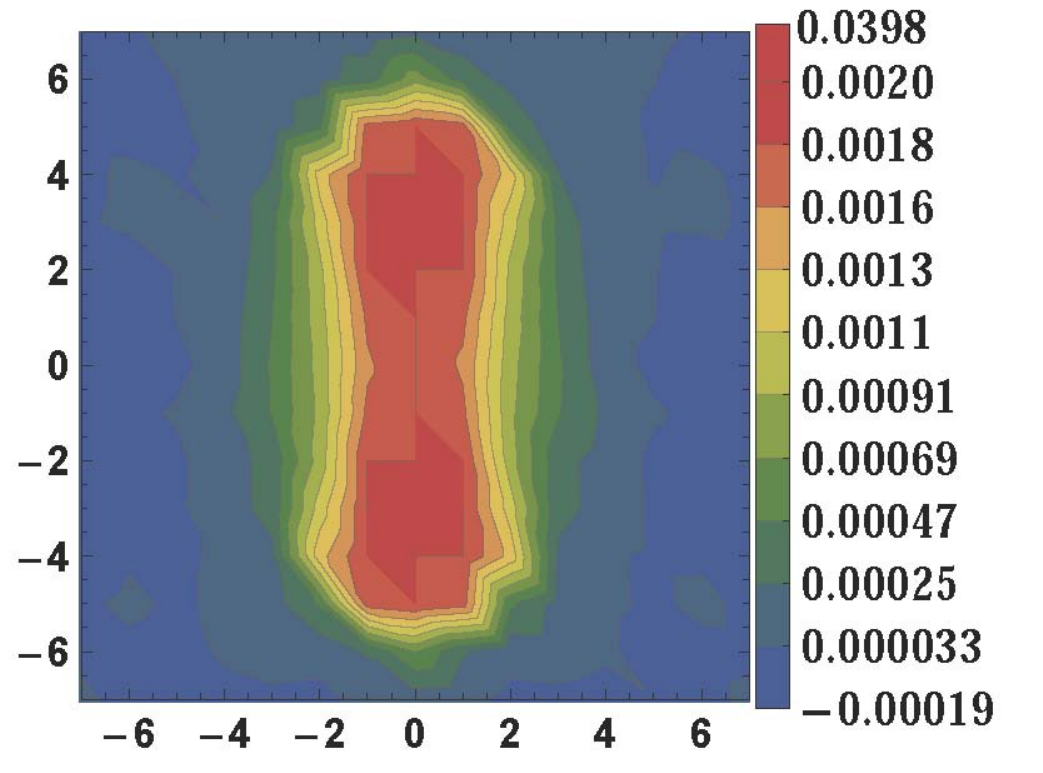}
\par\end{centering}}
    \subfloat[Chromomagnetic Field\label{fig:cfield_L_r1_0_r2_8_B}]{
\begin{centering}
    \includegraphics[width=3.5cm]{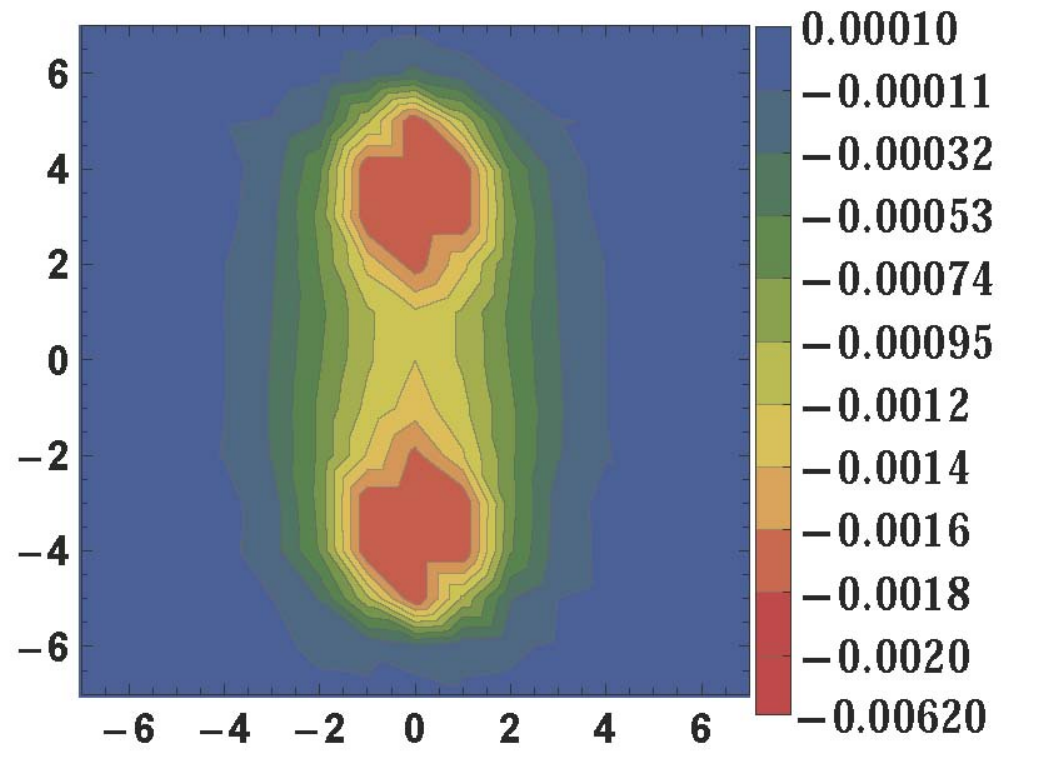}
\par\end{centering}}
    \subfloat[Energy Density\label{fig:cfield_L_r1_0_r2_8_Energ}]{
\begin{centering}
    \includegraphics[width=3.5cm]{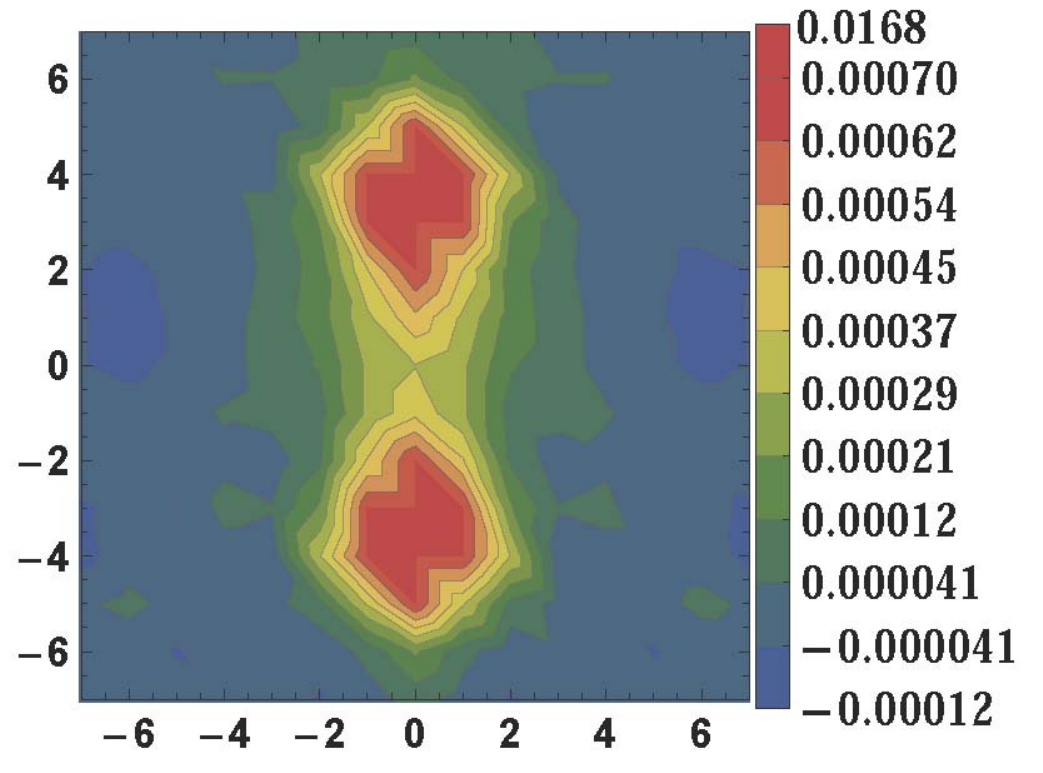}
\par\end{centering}}
    \subfloat[Action Density\label{fig:cfield_L_r1_0_r2_8_Act}]{
\begin{centering}
    \includegraphics[width=3.5cm]{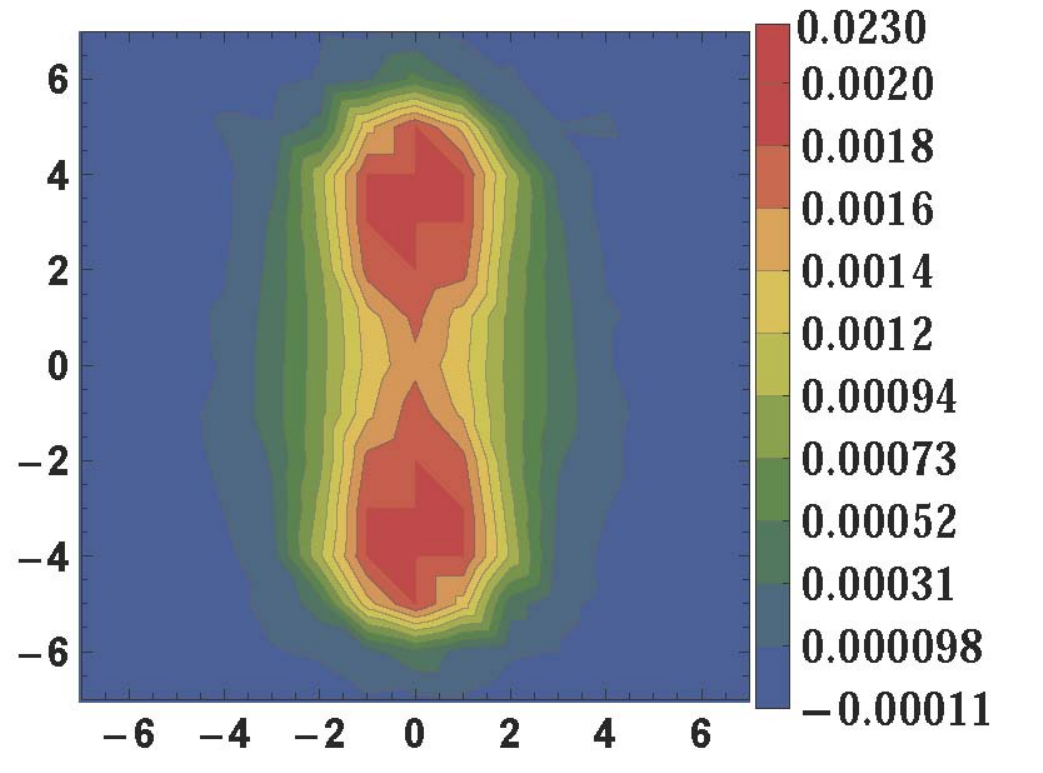}
\par\end{centering}}
\par\end{centering}
    \caption{Results for the L geometry with $r_1=0$ and $r_2=8$. The gluon and the antiquark
are superposed and the results are consistent with the static quark-antiquark case.}
    \label{latfig/cfield_L_r1_0_r2_8}
\end{figure}

\begin{figure}[H]
\begin{centering}
    \subfloat[Chromoelectric Field\label{fig:cfield_L_r1_2_r2_8_E}]{
\begin{centering}
    \includegraphics[width=3.5cm]{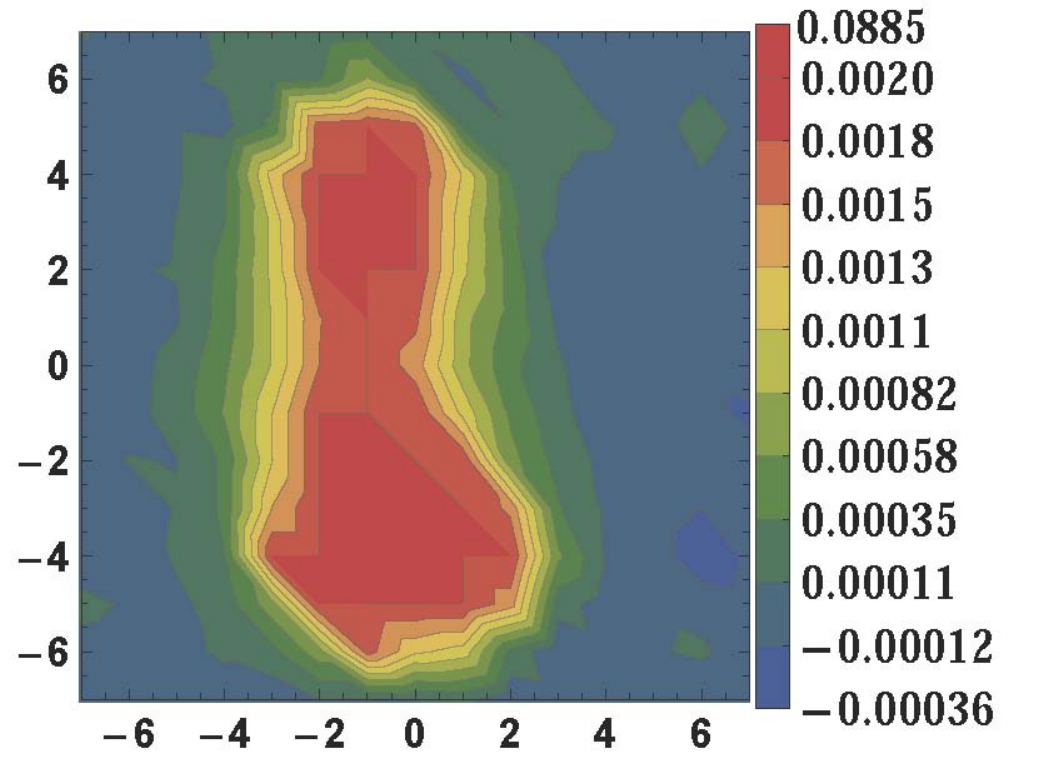}
\par\end{centering}}
    \subfloat[Chromomagnetic Field\label{fig:cfield_L_r1_2_r2_8_B}]{
\begin{centering}
    \includegraphics[width=3.5cm]{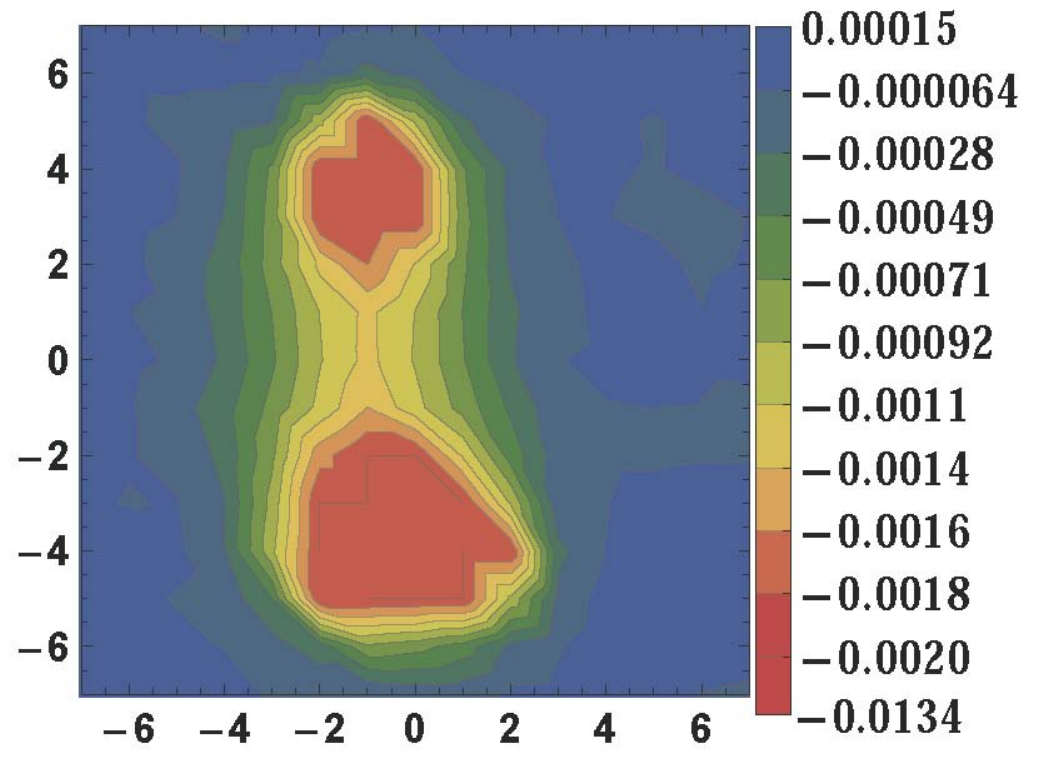}
\par\end{centering}}
    \subfloat[Energy Density\label{fig:cfield_L_r1_2_r2_8_Energ}]{
\begin{centering}
    \includegraphics[width=3.5cm]{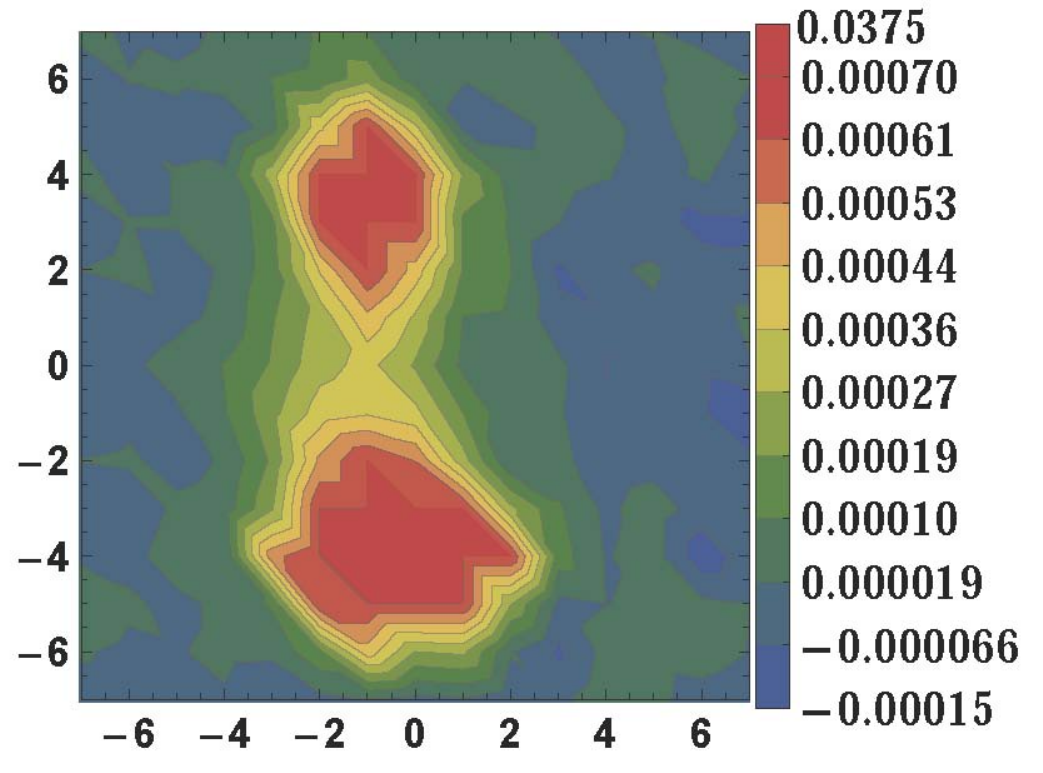}
\par\end{centering}}
    \subfloat[Action Density\label{fig:cfield_L_r1_2_r2_8_Act}]{
\begin{centering}
    \includegraphics[width=3.5cm]{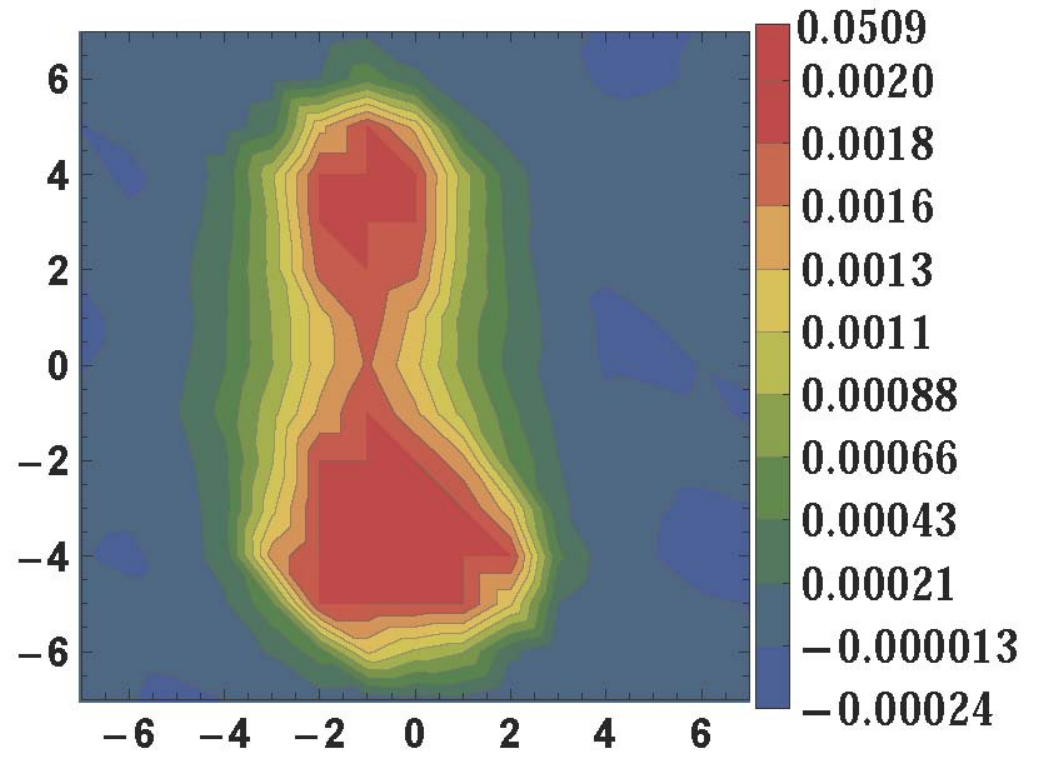}
\par\end{centering}}
\par\end{centering}
    \caption{Results for the L geometry with $r_1=2$ and $r_2=8$}
    \label{latfig/cfield_L_r1_2_r2_8}
\end{figure}

\begin{figure}[H]
\begin{centering}
    \subfloat[Chromoelectric Field\label{fig:cfield_L_r1_8_r2_8_E}]{
\begin{centering}
    \includegraphics[width=3.5cm]{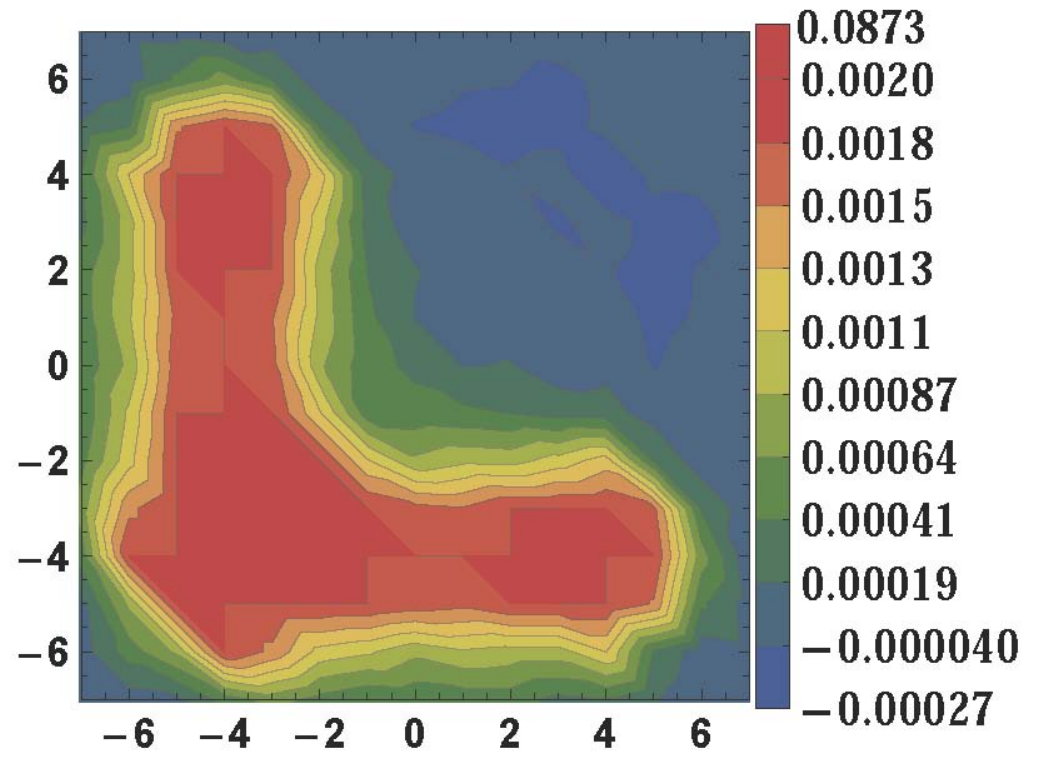}
\par\end{centering}}
    \subfloat[Chromomagnetic Field\label{fig:cfield_L_r1_8_r2_8_B}]{
\begin{centering}
    \includegraphics[width=3.5cm]{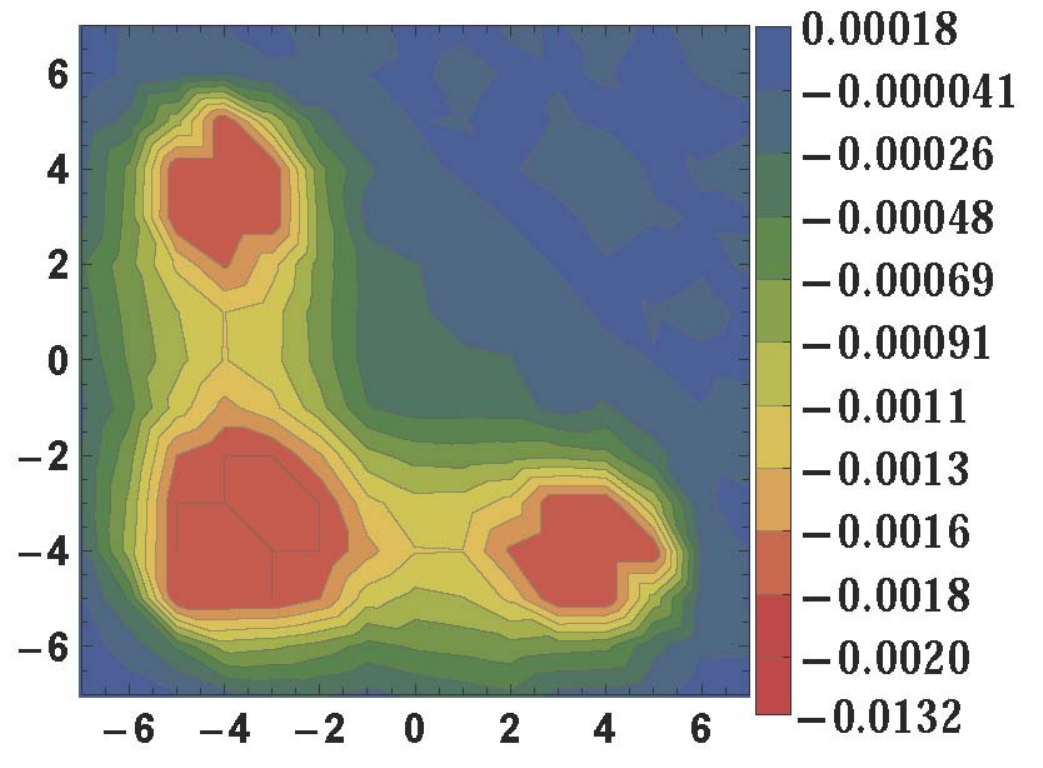}
\par\end{centering}}
    \subfloat[Energy Density\label{fig:cfield_L_r1_8_r2_8_Energ}]{
\begin{centering}
    \includegraphics[width=3.5cm]{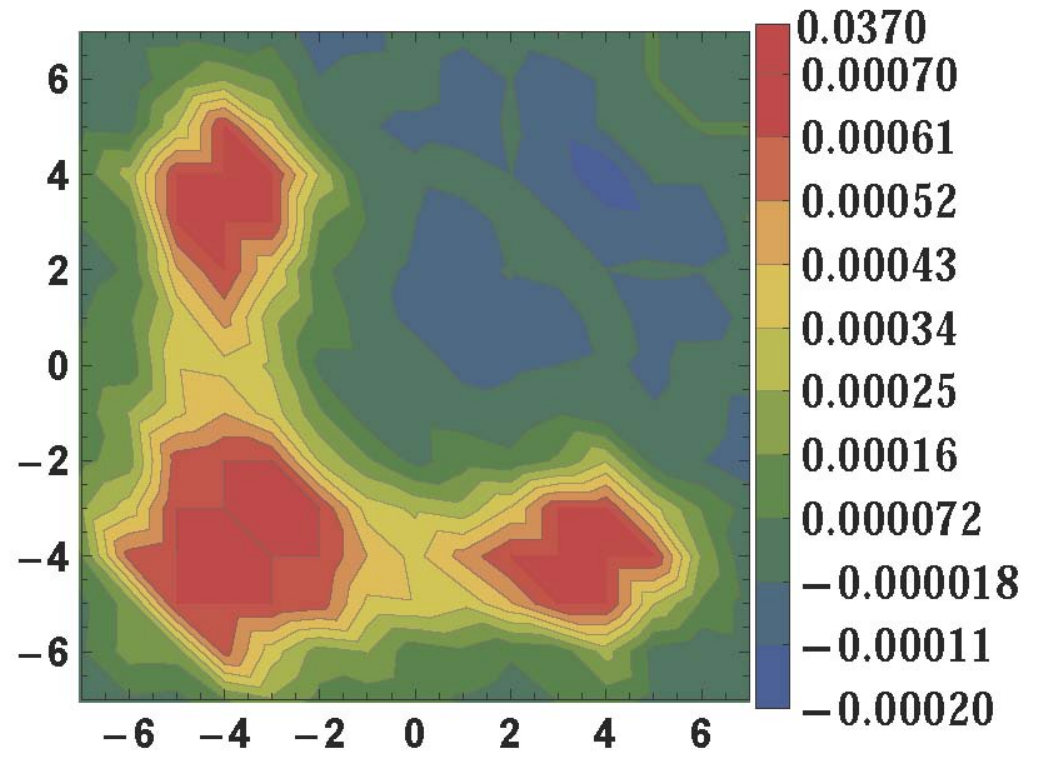}
\par\end{centering}}
    \subfloat[Action Density\label{fig:cfield_L_r1_8_r2_8_Act}]{
\begin{centering}
    \includegraphics[width=3.5cm]{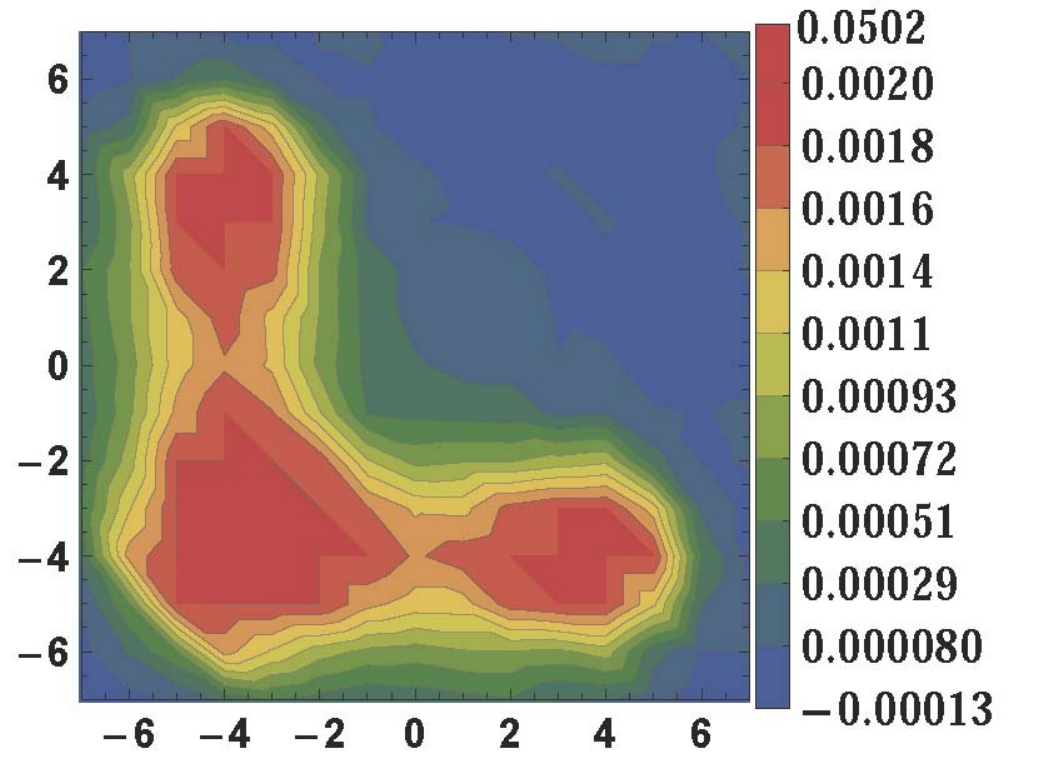}
\par\end{centering}}
\par\end{centering}
    \caption{Results for the L geometry with $r_1=8$ and $r_2=8$}
    \label{latfig/cfield_L_r1_8_r2_8}
\end{figure}

\begin{figure}[H]
\begin{centering}
    \subfloat[Chromoelectric Field\label{fig:cfield_L_r1_6_r2_6_E}]{
\begin{centering}
    \includegraphics[width=3.5cm]{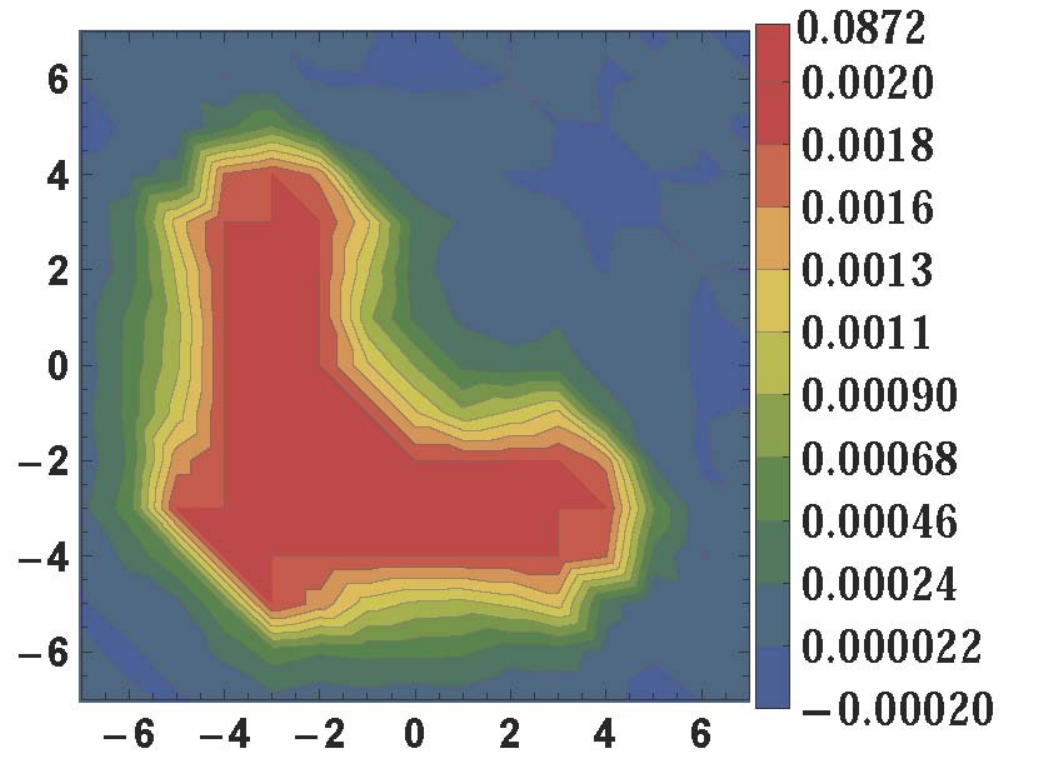}
\par\end{centering}}
    \subfloat[Chromomagnetic Field\label{fig:cfield_L_r1_6_r2_6_B}]{
\begin{centering}
    \includegraphics[width=3.5cm]{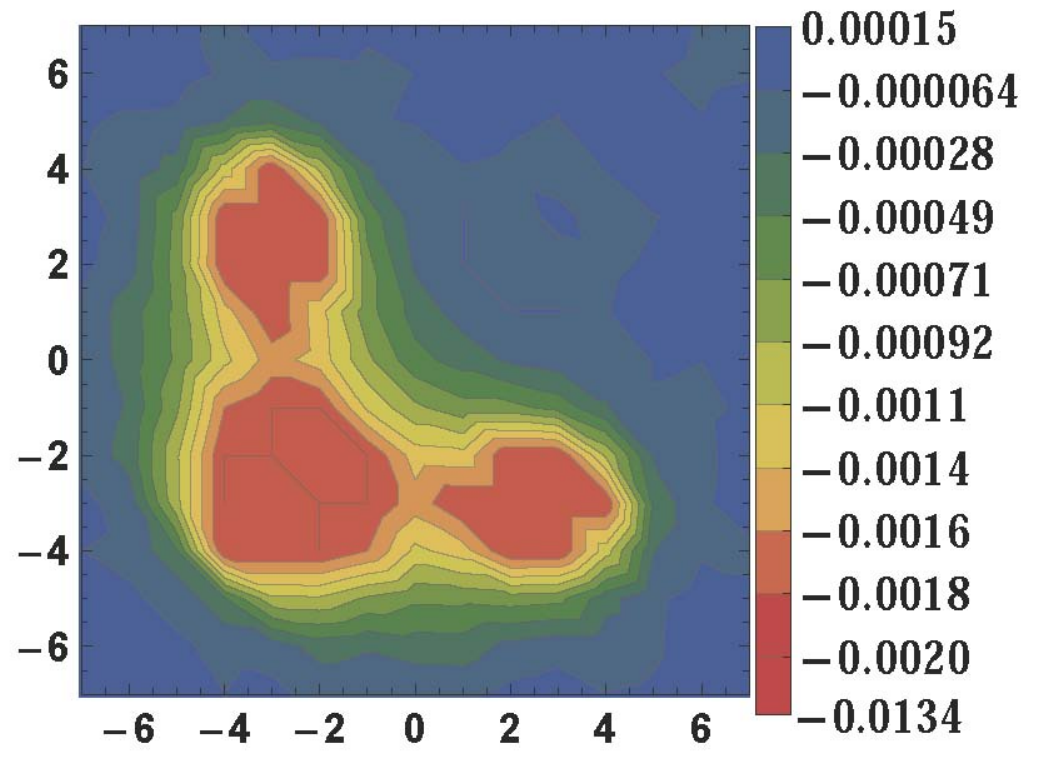}
\par\end{centering}}
    \subfloat[Energy Density\label{fig:cfield_L_r1_6_r2_6_Energ}]{
\begin{centering}
    \includegraphics[width=3.5cm]{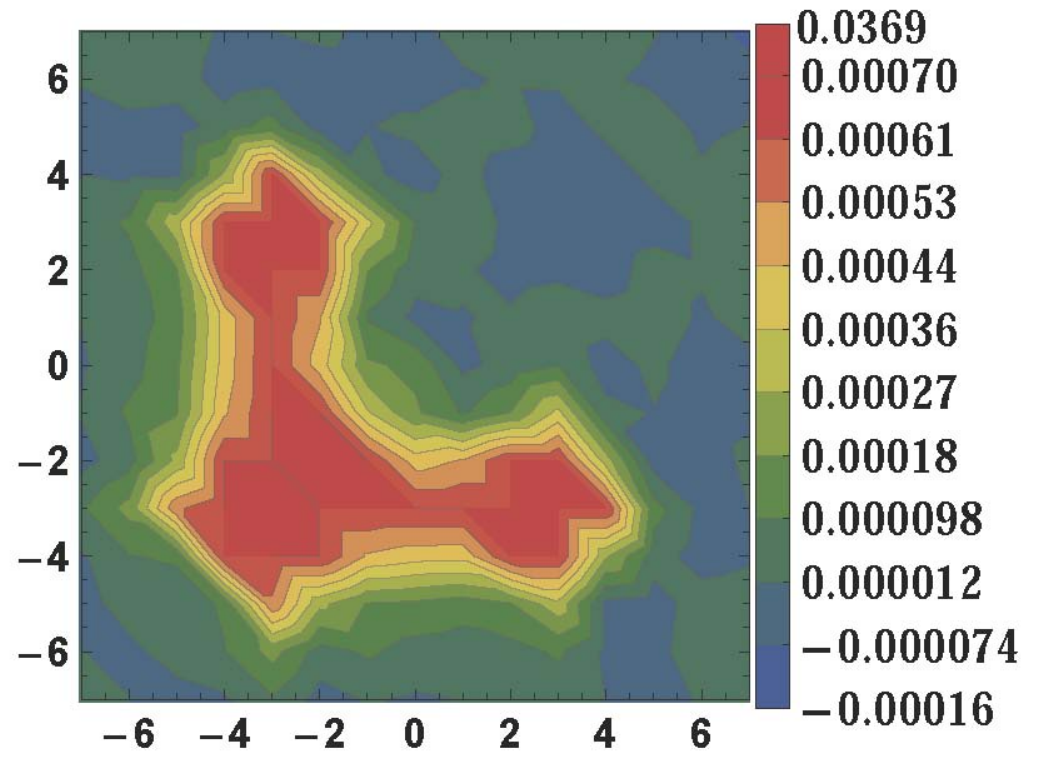}
\par\end{centering}}
    \subfloat[Action Density\label{fig:cfield_L_r1_6_r2_6_Act}]{
\begin{centering}
    \includegraphics[width=3.5cm]{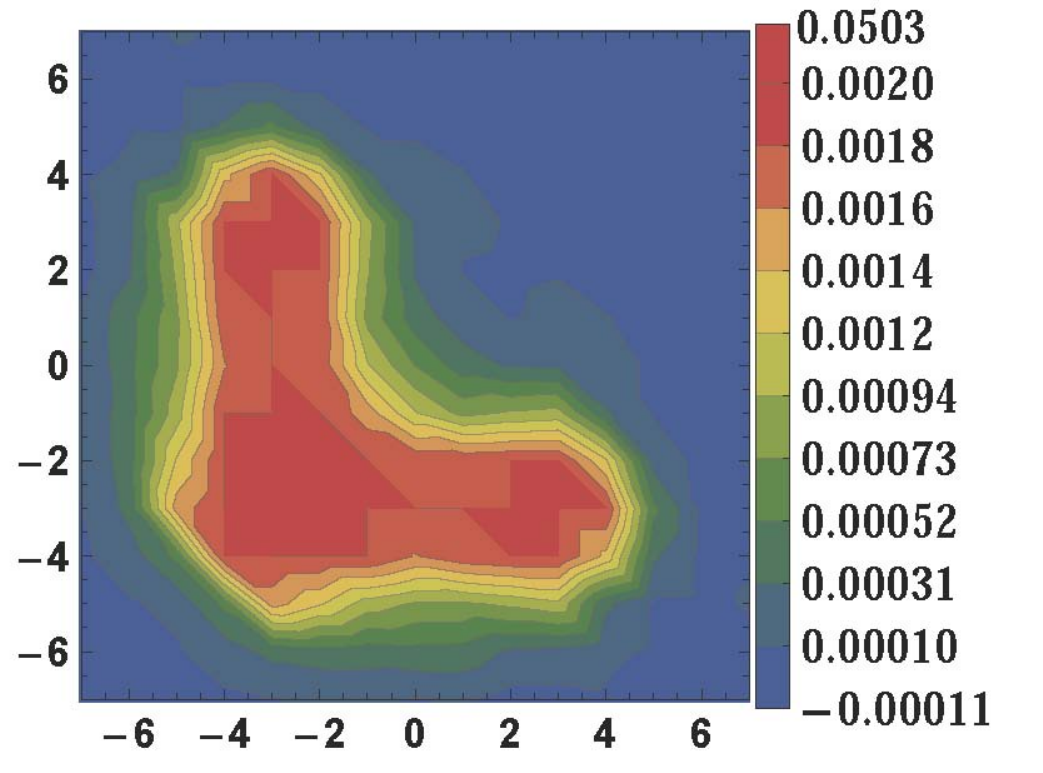}
\par\end{centering}}
\par\end{centering}
    \caption{Results for the L geometry with $r_1=6$ and $r_2=6$}
    \label{latfig/cfield_L_r1_6_r2_6}
\end{figure}

In the following pictures we present the results for the chromolectric and chromomagnetic fields and for the energy and action density. The figures \ref{latfig/cfield_U_d_0_l_8} to \ref{latfig/cfield_U_d_6_l_8} show the results for the U shape geometry and figures \ref{latfig/cfield_L_r1_0_r2_8} to \ref{latfig/cfield_L_r1_6_r2_6} the results for the L shape geometry.

\begin{figure}[H]
\begin{centering}

    \subfloat[\label{fig:cfield_U_d_0_l_8_All_2D}$d=0$ and $l=8$ for $x=0$]{
\begin{centering}
    \includegraphics[height=2.2cm]{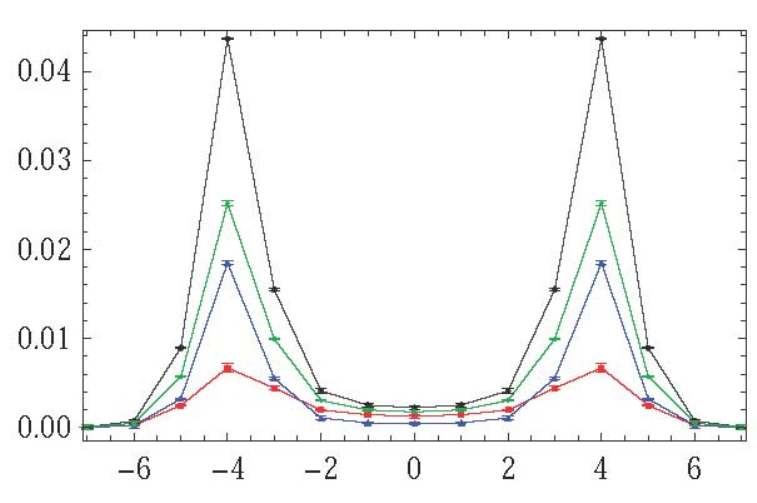}
\par\end{centering}}
    \subfloat[\label{fig:cfield_U_d_0_l_8_All0_2D}$d=0$ and $l=8$ for $y=4$]{
\begin{centering}
    \includegraphics[height=2.2cm]{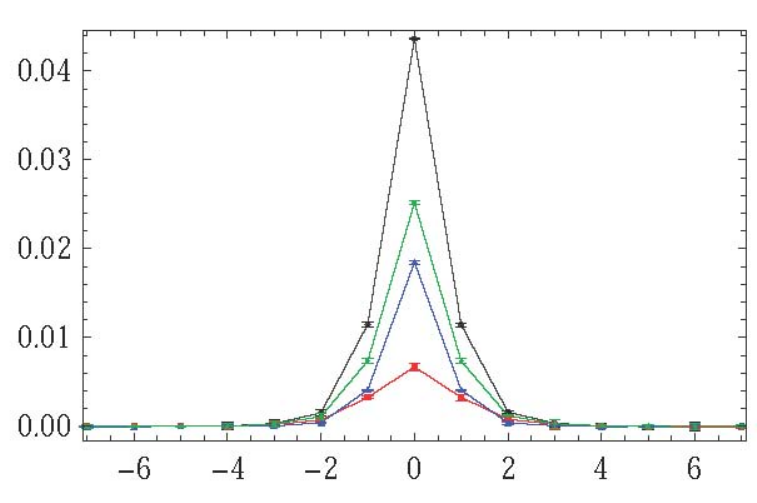}
\par\end{centering}}
    \subfloat[\label{fig:cfield_U_d_2_l_8_All_2D}$d=2$ and $l=8$ for $x=0$]{
\begin{centering}
    \includegraphics[height=2.2cm]{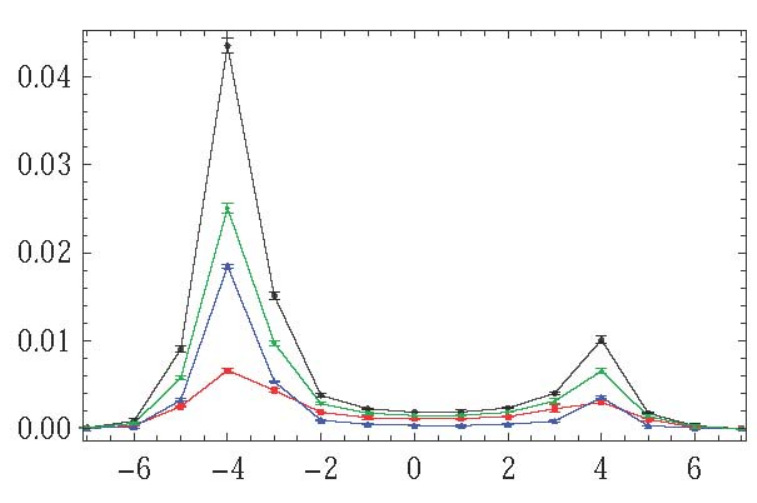}
\par\end{centering}}
    \subfloat[\label{fig:cfield_U_d_2_l_8_All0_2D}$d=2$ and $l=8$ for $y=4$]{
\begin{centering}
    \includegraphics[height=2.2cm]{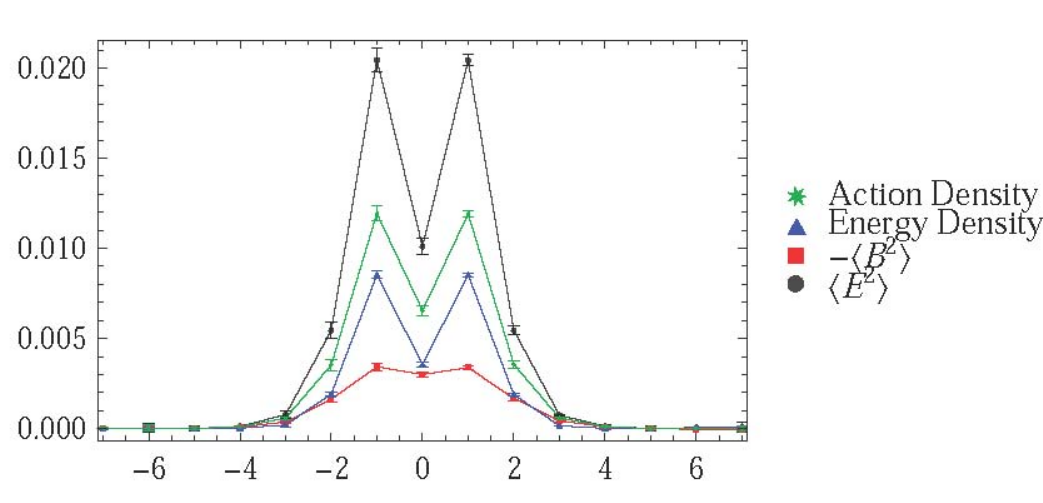}
\par\end{centering}}

    \subfloat[\label{fig:cfield_U_d_4_l_8_All_2D}$d=4$ and $l=8$ for $x=0$]{
\begin{centering}
    \includegraphics[height=2.2cm]{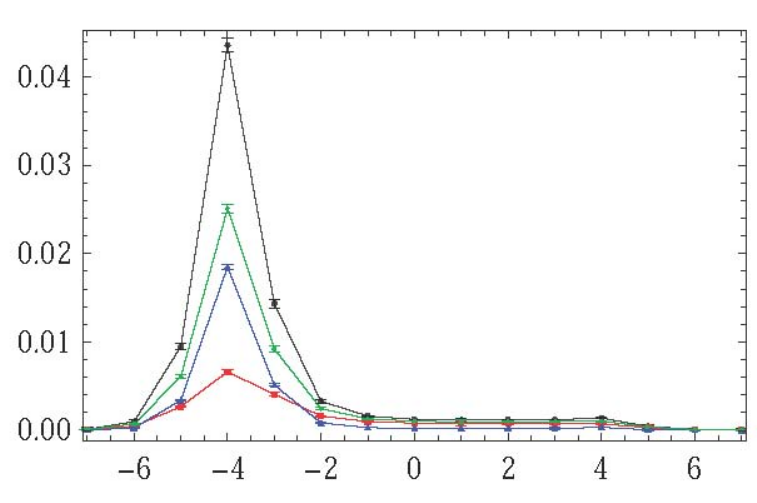}
\par\end{centering}}
    \subfloat[\label{fig:cfield_U_d_4_l_8_All0_2D}$d=4$ and $l=8$ for $y=4$]{
\begin{centering}
    \includegraphics[height=2.2cm]{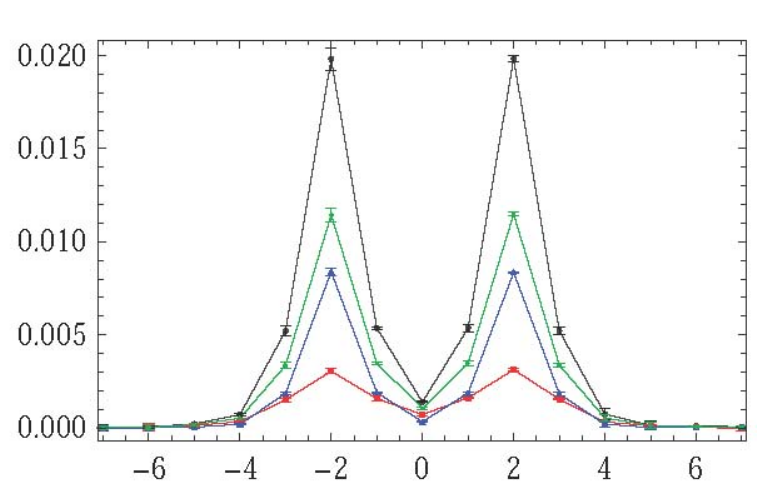}
\par\end{centering}}
    \subfloat[\label{fig:cfield_U_d_6_l_8_All_2D}$d=6$ and $l=8$ for $x=0$]{
\begin{centering}
    \includegraphics[height=2.2cm]{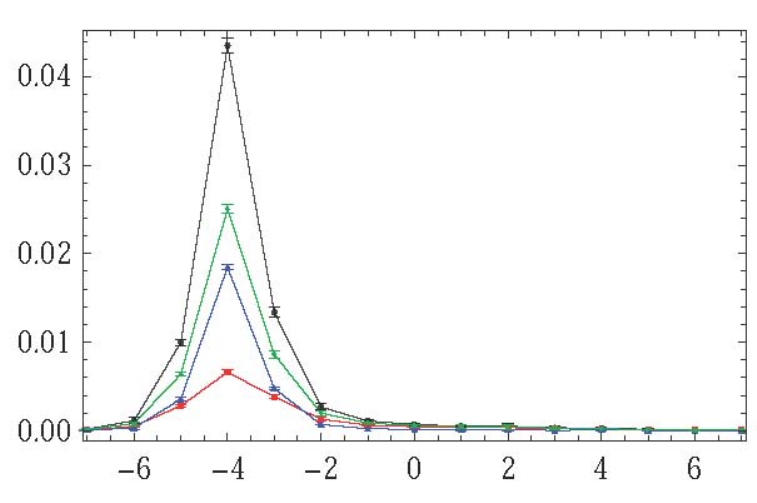}
\par\end{centering}}
    \subfloat[\label{fig:cfield_U_d_6_l_8_All0_2D}$d=6$ and $l=8$ for $y=4$]{
\begin{centering}
    \includegraphics[height=2.2cm]{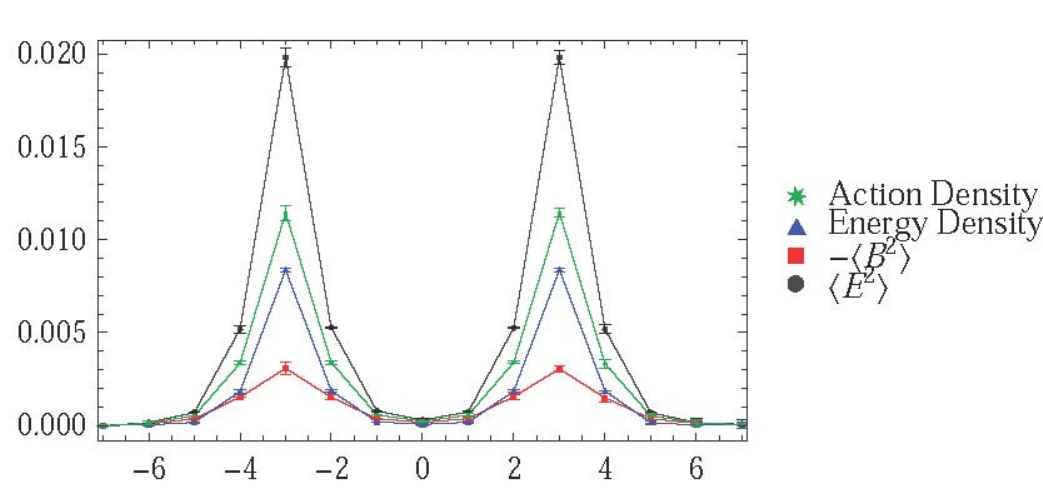}
\par\end{centering}}

\par\end{centering}

    \caption{Results for the U geometry.}
    \label{latfig/cfield_U_All}
\end{figure}

\begin{figure}[H]
\begin{centering}

    \subfloat[\label{fig:cfield_L_d_0_l_8_Allx_2D}$r_1=0$ and $r_2=8$, along segment gluon-antiquark]{
\begin{centering}
    \includegraphics[height=2.2cm]{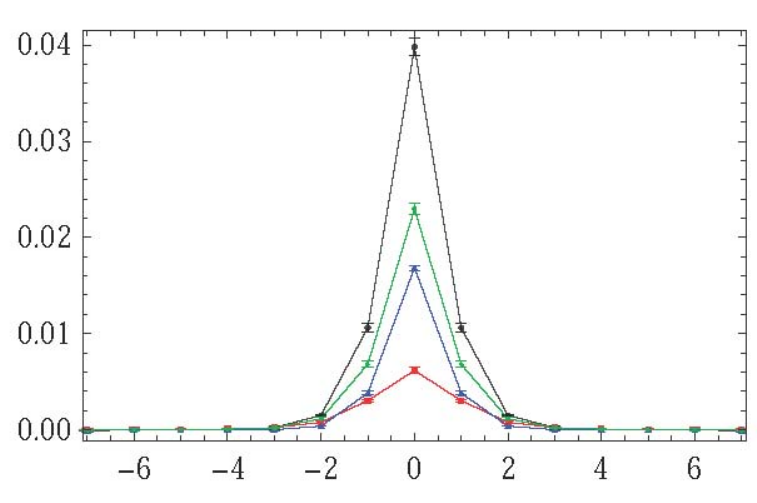}
\par\end{centering}}
    \subfloat[\label{fig:cfield_L_r1_0_r2_8_Ally_2D}$r_1=0$ and $r_2=8$, along segment gluon-quark]{
\begin{centering}
    \includegraphics[height=2.2cm]{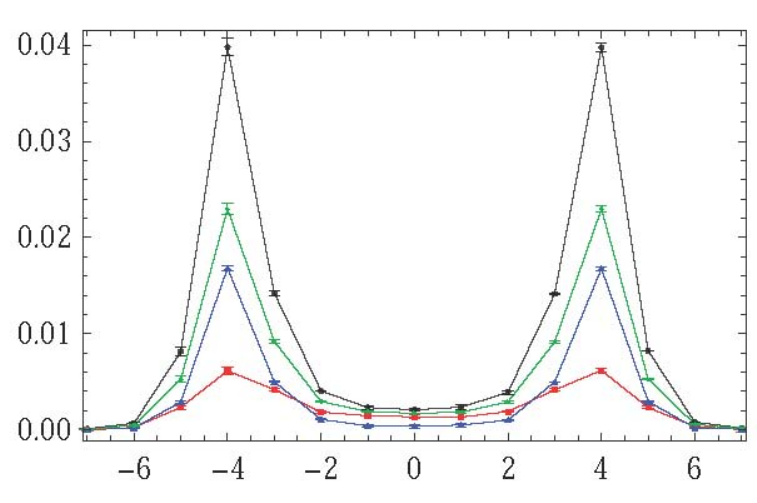}
\par\end{centering}}
    \subfloat[\label{fig:cfield_L_r1_2_r2_8_Allx_2D}$r_1=2$ and $r_2=8$, along segment gluon-antiquark]{
\begin{centering}
    \includegraphics[height=2.2cm]{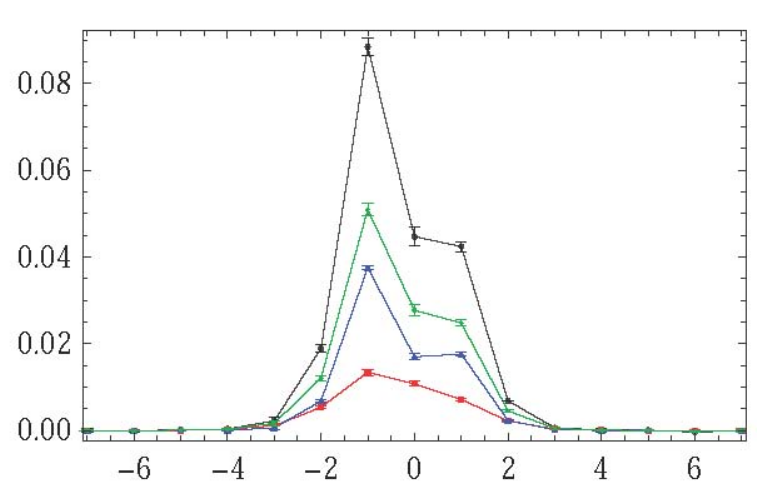}
\par\end{centering}}
    \subfloat[\label{fig:cfield_L_r1_2_r2_8_Ally_2D}$r_1=2$ and $r_2=8$, along segment gluon-quark]{
\begin{centering}
    \includegraphics[height=2.2cm]{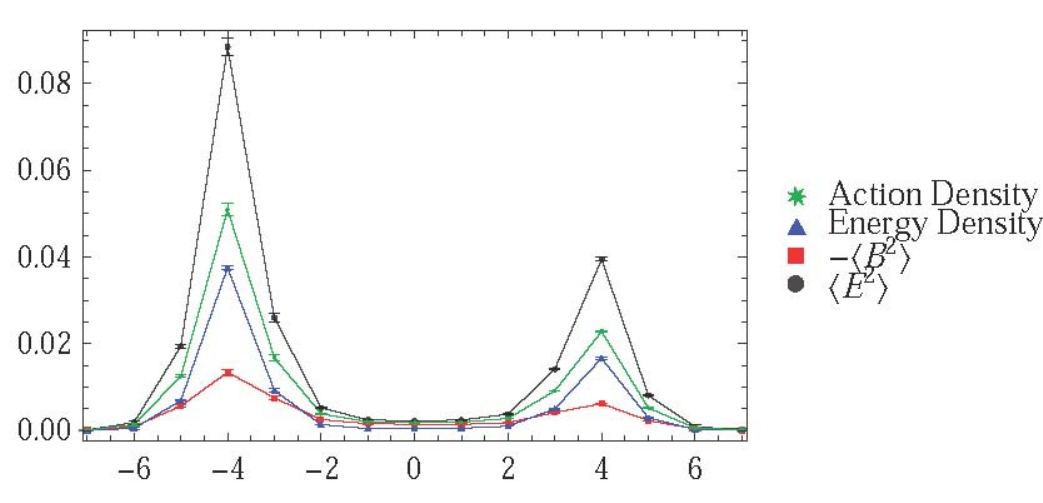}
\par\end{centering}}

    \subfloat[\label{fig:cfield_L_r1_8_r2_8_Allx_2D}$r_1=8$ and $r_2=8$, along segment gluon-antiquark]{
\begin{centering}
    \includegraphics[height=2.2cm]{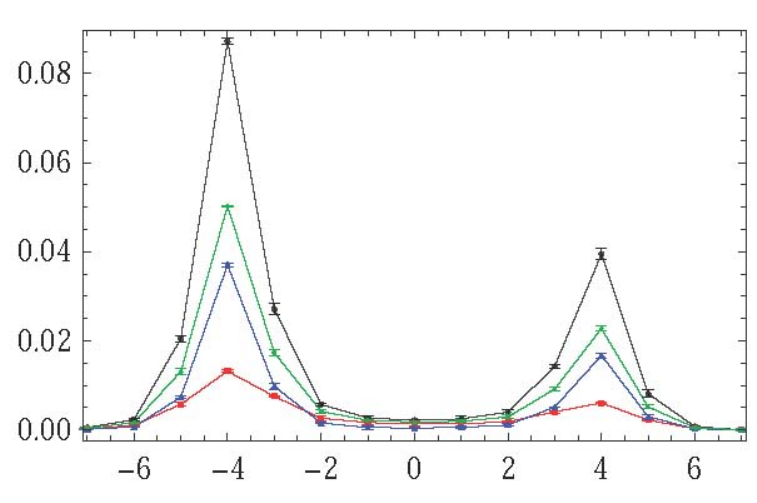}
\par\end{centering}}
    \subfloat[\label{fig:cfield_L_d_4_l_8_Ally_2D}$r_1=8$ and $r_2=8$, along segment gluon-quark]{
\begin{centering}
    \includegraphics[height=2.2cm]{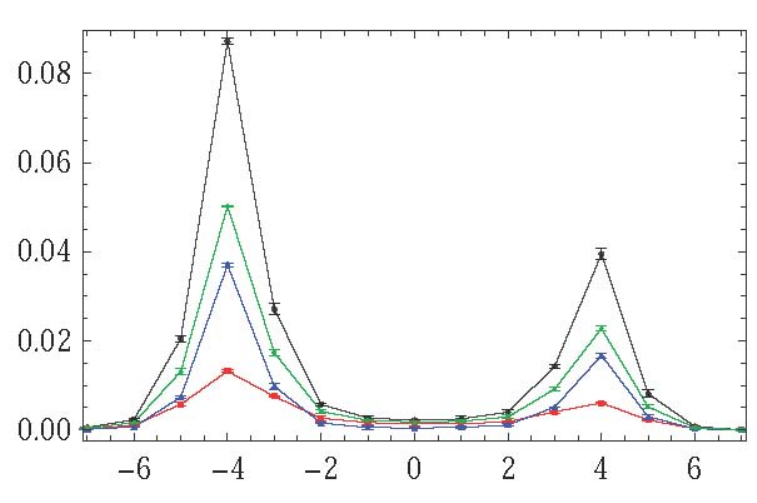}
\par\end{centering}}
    \subfloat[\label{fig:cfield_L_r1_6_r2_6_Allx_2D}$r_1=6$ and $r_2=6$, along segment gluon-antiquark]{
\begin{centering}
    \includegraphics[height=2.2cm]{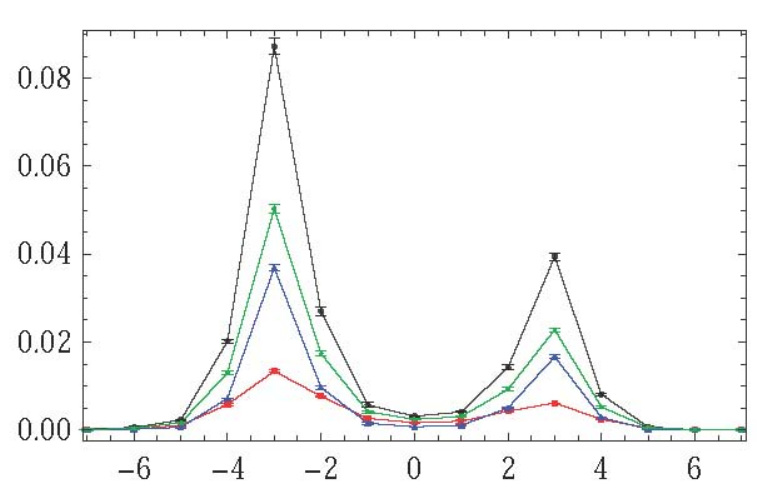}
\par\end{centering}}
    \subfloat[\label{fig:cfield_L_r1_6_r2_6_Ally_2D}$r_1=6$ and $r_2=6$, along segment gluon-quark]{
\begin{centering}
    \includegraphics[height=2.2cm]{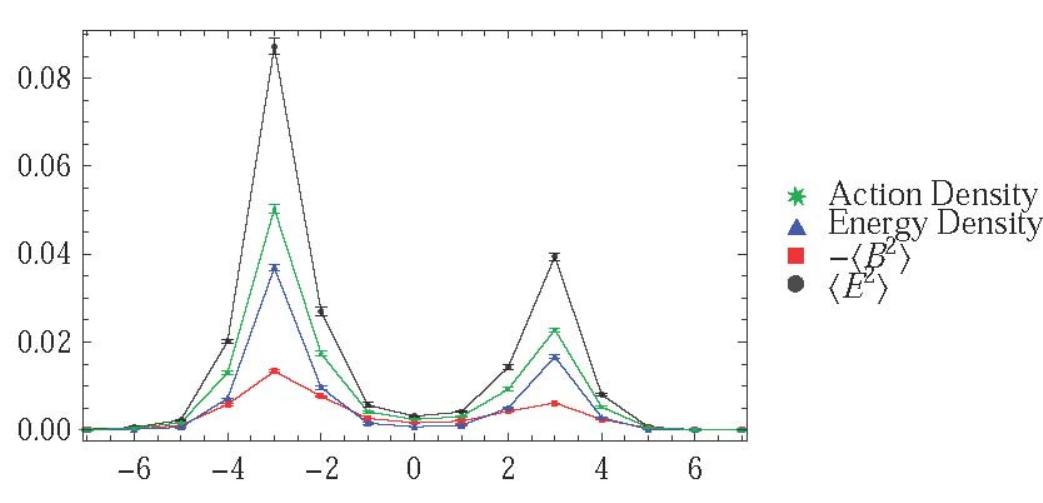}
\par\end{centering}}

\par\end{centering}

    \caption{Results for the L geometry.}
    \label{latfig/cfield_L_All}
\end{figure}

\begin{figure}[H]
\begin{center}
    \includegraphics[width=12cm]{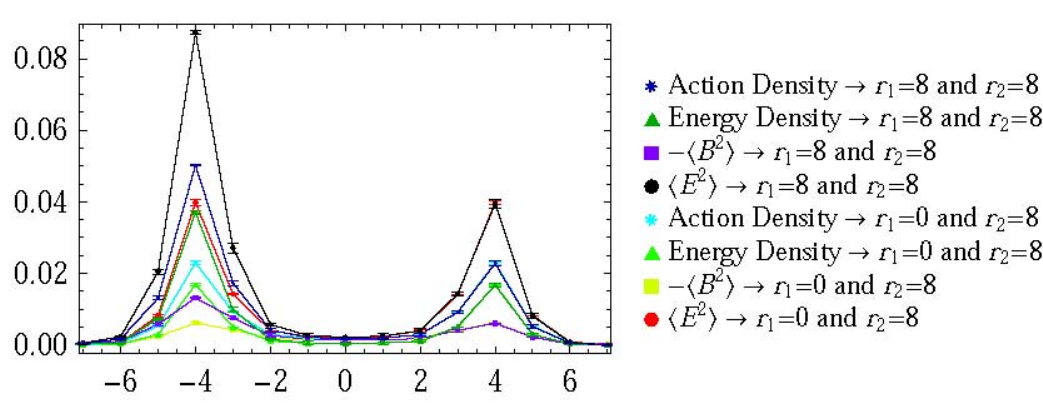}
    \caption{Comparison between the results for ($r_1 = 0; r_2 = 8$) and ($r_1 = 8; r_2 = 8$) along segment gluon-antiquark in the L geometry.}
    \label{qq_qqg}
\end{center}
\end{figure}

In figure \ref{latfig/cfield_U_All} we present the results for the U geometry along $x=0$, this corresponds to the axis along gluon and the middle distance between quark-antiquark, and $y=4$, this corresponds to the axis along quark-antiquark. This results show that when the quark and the antiquark are superposed the results are consistent with the degenerate case of the two gluon glueball.
In figure \ref{latfig/cfield_L_All} we present the results for the L geometry along the segment gluon-antiquark and along segment gluon-quark. And finally, in figure \ref{qq_qqg} we show a comparison between the results for $(r_1=0,r_2=8)$ and $(r_1=8,r_2=8)$ along segment gluon-antiquark in the L geometry, when the gluon and the antiquark are superposed this results are consistent with the static quark-antiquark case.

\section{A double vertex model for the scalar confining interaction}
\label{sec:potential}

Our lattice computation of flux tubes, string-like extended objects,
suggest that the confining potential is scalar since the groundstate of a bosonic string is a scalar. 
This is also suggested by the spectroscopy of hadrons.
In a pertubative QCD scenario, the hadron spectroscopy would be qualitatively 
similar to electronic spectra of the lighter atoms. 
It is remarkable that the Spin-Orbit potential (also called fine interaction 
in atomic physics) turns out to be suppressed in hadronic spectra because it is 
smaller than the Spin-Spin potential (also called hyperfine interaction).
This constitutes an evidence of non-pertubative QCD. Another evidence of
non-pertubative QCD is present in the angular and radial excitations of hadrons,
which fit linear trajectories in Regge plots, and suggest a long range,
probably linear, confining potential for the quarks. 
This led Henriques, Kellet and Moorhouse, Isgur and Karl, and others  
\cite{Henriques,Isgur}
to develop a Quark Model
where a short-range vector potential plus a long-range scalar potential 
partly cancel the Spin-Orbit interaction. The short range potential 
is inspired in the one gluon exchange, and the quark vertex is a Coulomb-like
potential, with a vector coupling $\bar \psi \gamma^\mu \psi$. The long range 
potential has  scalar coupling $\bar \psi \psi$, and is a linear potential.

On the other hand, the QCD Lagrangian is chiral invariant in the 
limit of vanishing quark masses, because the coupling of a gluon to a quark is
vector like. This is crucial because 
spontaneous chiral symmetry breaking is accepted to occur in 
low energy hadronic physics, for the light flavors $u$, $d$ and  $s$, where,
$m_u , m_d << m_s < \Lambda_{QCD} < M_N/3$.
The techniques of current algebra lead to 
beautifully correct theorems, the  PCAC (Partially Conserved
Axial Current) theorems. 
The Quark Models are widely used as a simplification of QCD, 
convenient to study quark bound states and hadron scattering. 
Recently 
\cite{Bicudo} 
we have shown that these beautiful PCAC theorems, 
like the Weinberg theorem for $\pi-\pi$ scattering, are reproduced by quark 
models with spontaneous chiral symmetry breaking.

\begin{figure}[h]
\includegraphics{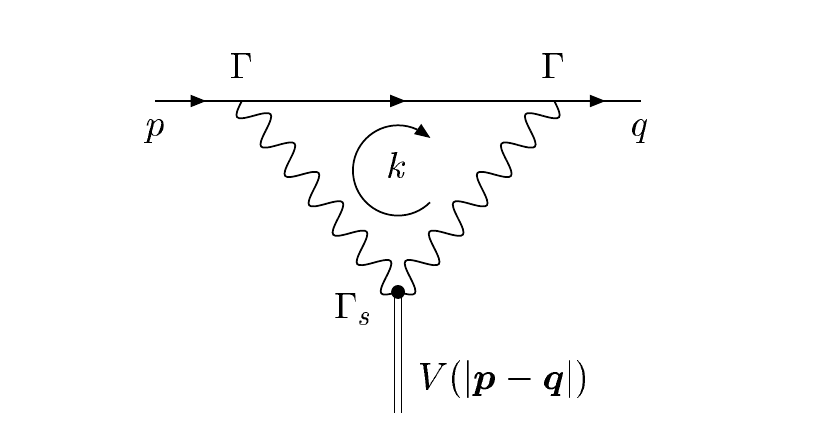}
\caption{The coupling of a quark to a string with a double gluon vertex} 
\label{loopvertex} 
\end{figure}

We aim to couple a quark line in a Feynman diagram with a 
scalar string, using the vector gluon-quark coupling of QCD. We remark that a simple 
vertex does not solve this problem, therefore we reformulate the standard 
coupling of the quark to the confining potential. The coupling needs at least a 
double vertex, similar to the vertices that couple a quark to a gluon ladder in 
models of the pomeron. Our double vertex is depicted in Fig. \ref{loopvertex}. 

To get the Dirac coupling of each gluon to a light quark,
we follow the coupling obtained in the heavy-light quark system, computed
in the local coordinate gauge
\cite{Nora}. This results in a Dirac coupling a with pair of 
$\gamma^0$ matrices, which is also compatible with the Coulomb gauge.

In the color sector, each sub-vertex couples with the Gell-Mann 
matrix $\lambda^a /2$. Moreover the string is also a colored object because
it contains the flux of color electric field. In quark models the string usually
couples with a $\lambda^a /2$ to the quark line, here it couples with the
two $\lambda^a /2$ of the sub-vertices. For a scalar coupling, which is
symmetric, we use the symmetric structure function $d^{abc}$ defined with,
\begin{equation}
\{ \lambda^a ,\lambda^b \}= d^{abc}\lambda_c \ . 
\end{equation}

\begin{figure}[h]
\includegraphics{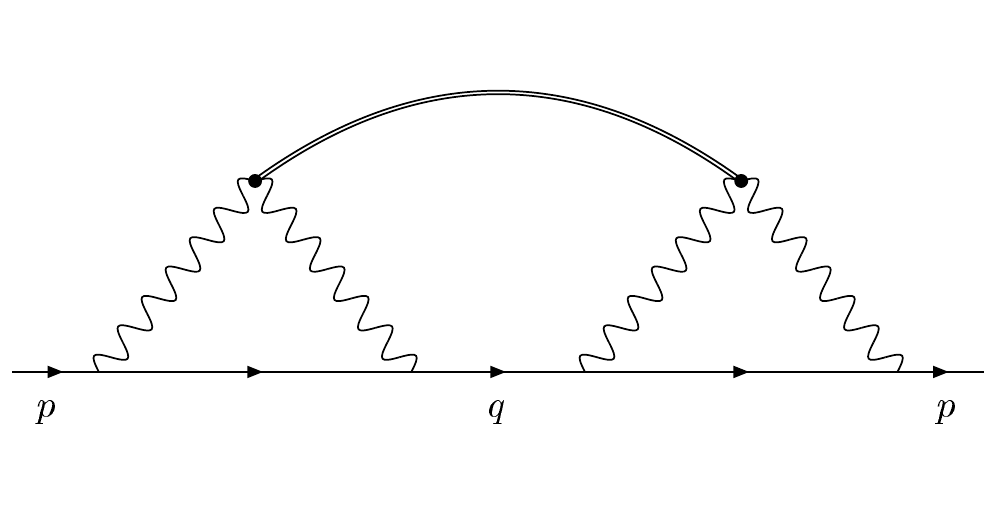}
\caption{ The self-energy term of the mass gap equation} 
\label{massgapfig} 
\end{figure}

The dependence in the relative momentum must comply with the linear confinement which
is derived from the string picture,
\begin{equation}
-i \, V( {\bf p}-{\bf q} )=-i \, { \sigma \over {\cal C}}{ - 8 \pi \over |{\bf p}-{\bf q} |^4}
\end{equation}
where $\sigma\simeq 200MeV/Fm$. is the string constant, and ${\cal C}$ is an algebraic color 
factor.

The gluon propagators and the different sub-vertices result in a distribution in the loop
momentum $k$, see Fig. \ref{loopvertex}. Here different choices would be possible. For simplicity 
we assume that the same ${p-q \over 2}$ momentum flows in each of the two effective gluon lines. 
We also remark that the distribution in $k$ is normalized to unity once the correct string tension 
is included in the relative potential $V(p-q)$. This amounts to consider that the momentum
$k$ distribution is a Dirac delta,
\begin{equation}
(2 \pi)^3 \delta^3 \left({\bf k} -{ {\bf p} + {\bf q} \over 2}\right) \ .
\end{equation}

We finally compute the vertex, decomposing the Dirac propagator in the convenient 
particle and anti-particle propagators, computing the energy loop integral, and 
summing in color indices,
\begin{equation}
{\cal V}_{eff} = \lambda^c \ 
(S_k + C_k\, \vers{k}\cdot\trD{\gamma}) \ 
\Biggr|_{k={p+q \over 2}} \ , 
\label{vertex result}
\end{equation}
where $S_k=m_k/\sqrt{k^2+{m_k}^2}, \
C_k=k/\sqrt{k^2+{m_k}^2}$, 
and where $m_k$ is the constituent quark mass, to be determined in the next section.

\section{Results for the model of scalar confinement}
\label{sec:conclusion}

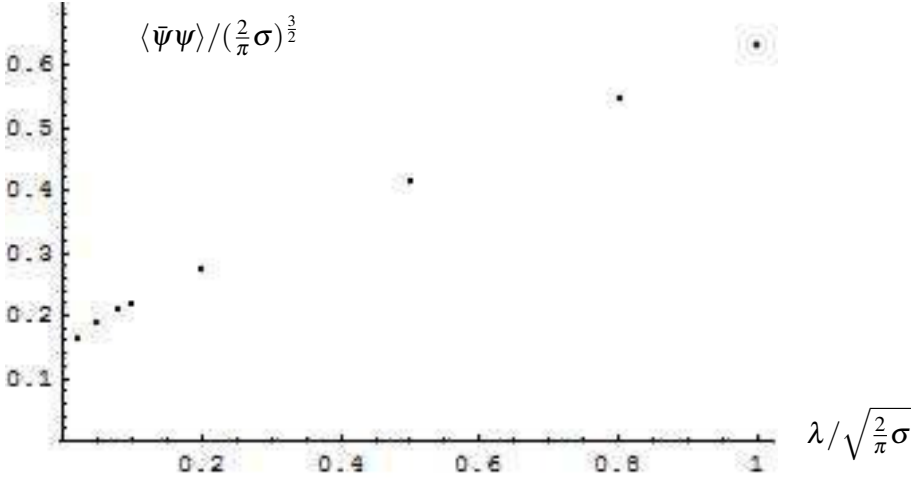
\begin{figure}[h]
\includegraphics[angle=-90]{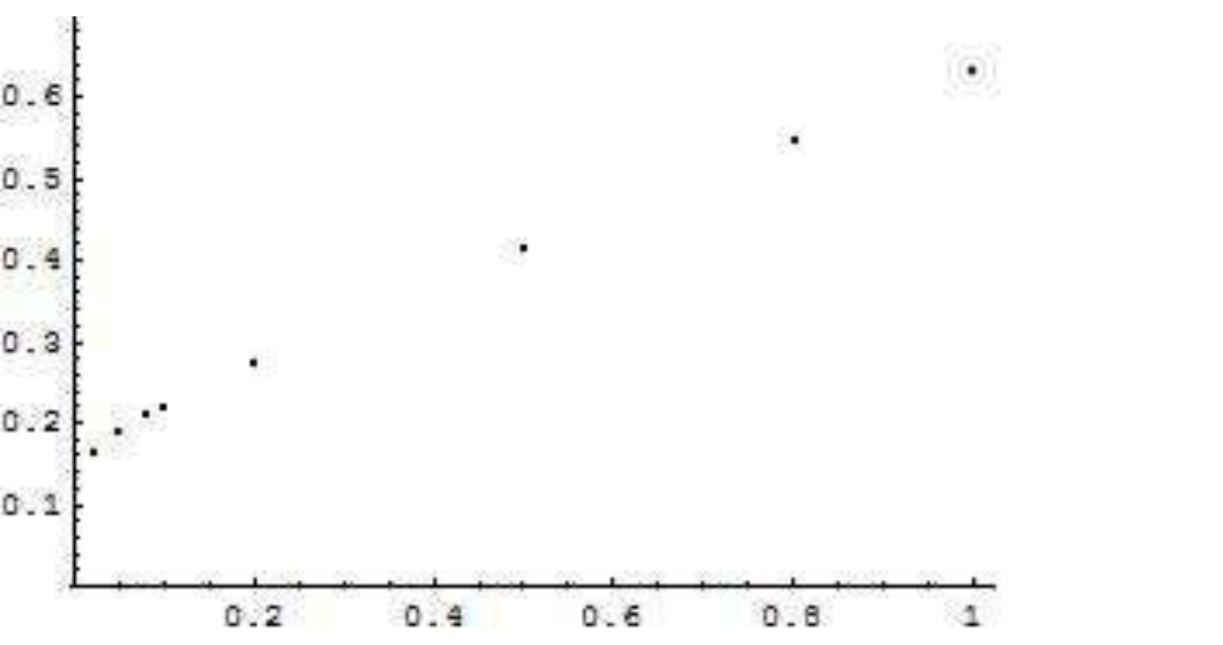}
\caption{ Testing the convergence of the numerical method with the quark
condensate $\langle \bar \psi \psi \rangle$.} 
\label{testing in condensate} 
\begin{picture}(0,0)(0,0)
\put(50,200){$\langle \bar \psi \psi \rangle /({2\over \pi}\sigma)^{3\over 2}$}
\put(300,50){$\lambda /\sqrt{{2\over \pi}\sigma}$}
\end{picture}
\end{figure}

We derive the mass gap equation projecting the Schwinger-Dyson equation with Dirac spinors,
\begin{equation}
\spub{s}{p}{\cal S}_0^{-1}(p)\spv{s'}{p} - \spub{s}{p}\Sigma(p)\spv{s'}{p} =0
\label{mass gap eq}
\end{equation}
where ${\cal S}_0$ is the free Dirac propagator, and where $\Sigma(p)$ 
is the self-energy depicted in Fig. \ref{massgapfig}. The self energy consits
in a three loop Feynman diagram, including one loop in each double vertex and
a third rainbow-like loop for the string exchange interaction.
Nevertheless each double vertex is simple, see eq. (\ref{vertex result}), and we get
for the self-energy term, 
\begin{eqnarray}
\spub{s}{p}\Sigma(p)\spv{s'}{p} &=& \hspace{3mm}\int\frac{d^3q}{(2\pi)^3}\left[
(C_k^2-S_k^2)S_q C_p + 2S_k C_k S_q S_p\vers{k}\cdot\vers{p} - \right. 
\nonumber\\ 
&& - C_q S_p\vers{q}\cdot\vers{p} - 2S_k C_k C_q C_p \vers{k}\cdot\vers{q} + 
 \left. + 2C_k^2 C_q S_p \vers{k}\cdot\vers{q}\ \vers{k}\cdot\vers{p}
\right]V\left(|\trD{p}-\trD{q}|\right) \ .
\label{self energy eq}
\end{eqnarray}

The mass gap equation is a difficult non-linear integral equation,
that does not converge with the usual methods
\cite{Yaouanc,Adler,Bicudo.solutions}. Here we develop a method to solve
it with a differential equation, using a convergence parameter 
$\lambda \rightarrow 0$. 
Our technique consists in starting with a large infrared cutoff $\lambda$,
where the integral term in eq. (\ref{self energy eq}) is small. Then
eq. (\ref{mass gap eq}) for the chiral angle $\varphi_p$ is essentially 
a differential equation which can be solved with the standard shooting method
\cite{Bicudo.solutions}. 
Next we decrease step by step the $\lambda$ parameter, using as an
initial guess for the evaluation of the integral the $\varphi_p$ determined
for the previous value of $\lambda$. 
We test the convergence of the 
method computing the quark condensate $\langle \bar \psi \psi \rangle $,
see Fig. \ref{testing in condensate}.

\vspace{1cm}
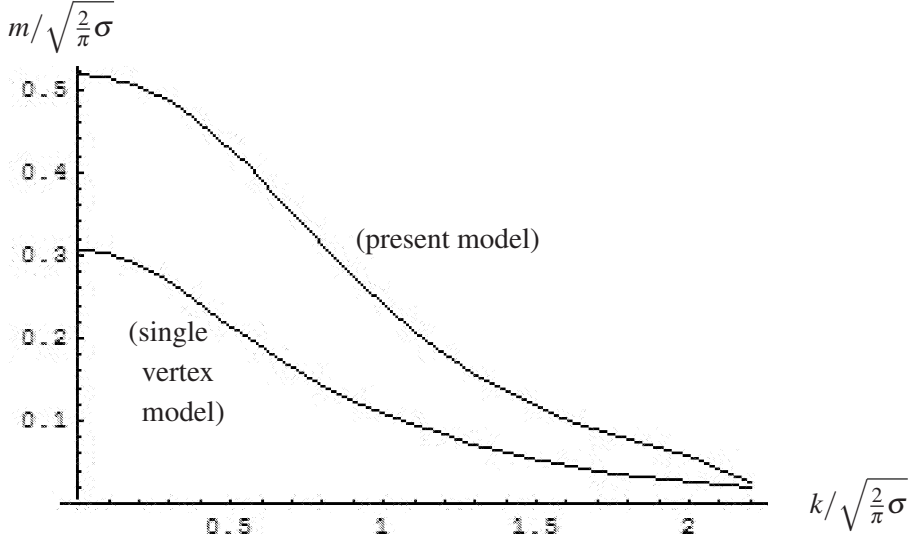
\begin{figure}[h]
\includegraphics[angle=-90]{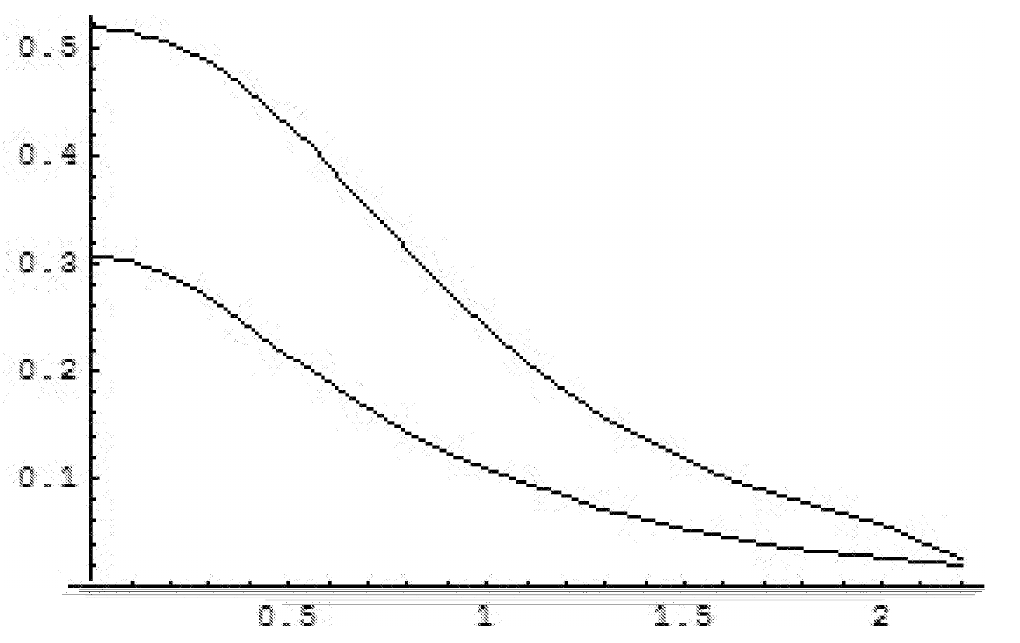}
\caption{ The $m_k$ solutions of the mass gap equation in units of 
$\sqrt{{2\over \pi}\sigma}$.} 
\label{massgapsolved}
\begin{picture}(0,0)(0,0)
\put(130,150){(present model)}
\put(45,115){(single}
\put(50,100){vertex}
\put(50,85){model)}
\put(300,50){$k/\sqrt{{2\over \pi}\sigma}$}
\put(00,230){$m/\sqrt{{2\over \pi}\sigma}$}
\end{picture}
\end{figure}

We solve the mass gap equation for the spontaneous breaking of chiral symmetry.
In Fig. \ref{massgapsolved} we compare the constituent quark mass $m_k$, computed
with our double vertex defined in eq. (\ref{vertex result}), with the mass computed
with a simple Coulomb gauge 
\cite{Adler} 
vertex $\lambda^c \, \gamma^0$. It turns out that
the dynamical quark mass $m_k$ computed with the double vertex is closer to the one
estimated is QCD sum rules, and this is a good point for the present work.

In the chiral limit of a vanishing quark mass, the effective vertex 
(\ref{vertex result}) 
${\cal V}_{eff}\rightarrow \lambda^c \vers{k}\cdot\trD{\gamma}$ is
proportional to $\gamma^\mu$ and is therefore chiral invariant as it should be, 
whereas in the heavy quark limit
${\cal V}_{eff}\rightarrow \lambda^c$ is simply a scalar vertex.
The dynamical generation of a quark 
mass $m_k$ also generates a scalar coupling for light quarks.

\section{Conclusions}

We investigate the components of the chromoelectric and chromomagnetic fields around a static gluon-quark-antiquark
\cite{Cardoso:2009qt}. 
We study the fields with two geometries, U and L shaped for the Wilson loops. This allows us to determine the shape of the flux-tubes with respect both to the energy and action densities. When the quark and the antiquark are superposed the results are consistent with the degenerate case of the two gluon glueball, and when the gluon and the antiquark are superposed the results are consistent with the static quark-antiquark case.
The general shape of the flux-tubes observed here reinforce the conclusions of Bicudo et al. \cite{Bicudo2008a} and Cardoso et al. 
\cite{Cardoso2007}.
Another interesting result is that the absolute value of the chromoelectric field dominates over the absolute value of the chromomagnetic field.

Moreover, in this talk we also present a model for the coupling of a quark to a scalar string
\cite{Bicudo:2003rw}. 
Since the string is composed of bosonic gluons, it is expected that the groundstate of the string, dominating confinement, is a scalar. To model scalar coupling, double vector
vertices are used, and the quark confining interaction has a single parameter, the
string tension $\sigma\simeq 200MeV/Fm$.
The results are encouraging, our model not only enhances chiral symmetry breaking, but the matching from vector confinement 
at quark high momenta and scalar confinement at small quark momenta is beautifully achieved. 
A similar matching has been obtained by Alkofer, Fischer, Llanes-Estrada and Schwenzer
\cite{Alkofer:2008tt}.
We will try in the future to reproduce the whole hadron 
spectrum.

\acknowledgments
This work was financed by the FCT contracts POCI/FP/81933/2007 and CERN/FP/83582/2008.


\end{document}